  \documentclass{aa}

\usepackage{graphicx}
\usepackage{txfonts}
\usepackage{xcolor} 

\usepackage{natbib}
\bibpunct{(}{)}{;}{a}{}{,} 

\begin{document}

   \title{A high redshift population of galaxies at the North Ecliptic Pole}

   \subtitle{unveiling the main sequence of dusty galaxies. }

   \author{L. Barrufet \inst{1,2}
          \and
          C. Pearson \inst{2,3,4}
          \and
          S. Serjeant \inst{4}
          \and
          K. Ma\l ek \inst{5,6}
          \and
          I. Baronchelli \inst{7}
          \and
          M.C. Campos-Varillas \inst{2,8}
           \and         
          G.J. White \inst{2,4}
          \and
          I. Valtchanov \inst{9}
          \and
          H. Matsuhara \inst{10}
          \and
          L. Conversi \inst{1}
          \and
          S. J. Kim \inst{11}
          \and
          T. Goto \inst{11}
          \and
          N. Oi \inst{12}
          \and
          M. Malkan \inst{13}
          \and
          H. Kim \inst{13}
          \and
          H. Ikeda \inst{14}
          \and 
          T.Takagi \inst{15}
          \and 
          Y. Toba \inst{16, 17, 18}
         \and 
          T. Miyaji \inst{19}
                   } 
   \institute{ European Space Astronomy Center, 28691 Villanueva de la Ca\~nada, Spain  \\    
	  \email{lbarrufet@sciops.esa.int}
         \and
         RAL Space, STFC Rutherford Appleton Laboratory, Didcot, Oxfordshire, OX11 0QX, UK  
         \and   
            Oxford Astrophysics, University of Oxford, Keble Rd, Oxford OX1 3RH, UK         
         \and
           Department of Physical Sciences, The Open University, Milton Keynes, MK7 6AA, UK 
         \and
          National Centre for Nuclear Research ul. Pasteura 7, 02-093 Warsaw, Poland 
         \and
         Aix Marseille Univ, CNRS, CNES, LAM Marseille, France   
         \and     
         Dipartimento di Fisica e Astronomia, Università di Padova, vicolo Osservatorio, 3, 35122 Padova, Italy. 
          \and     
         Astronomy Centre, Department of Physics and Astronomy, University of Sussex, Brighton BN1 9QH, UK 
        \and     
               Telespazio Vega UK for ESA, European Space Astronomy Centre, Operations Department, 28691 Villanueva de la Ca\~nada, Spain 
              \and     
        Institute of Space and Astronautical Science, Japan Aerospace Exploration Agency, 3-1-1 Yoshinodai, Chuo, Sagamihara, Kanagawa 252-5210, Japan  
        \and
        National Tsing Hua University, No. 101, Section 2, Kuang-Fu Road, Hsinchu, Taiwan 30013 
       \and
       Tokyo University of Science, 1-3 Kagurazaka, Shinjuku-ku, Tokyo 162-8601, Japan 
       \and
       Department of Physics and Astronomy, UCLA, Los Angeles, CA, 90095-1547, USA 
       \and 
       National Astronomical Observatory, 2-21-1 Osawa, Mitaka, Tokyo, Japan 
       \and
       Japan Space Forum, 3-2-1, Kandasurugadai, Chiyoda-ku, Tokyo 101-0062 Japan 
       \and
       Department of Astronomy, Kyoto University, Kitashirakawa-Oiwake-cho,Sakyo-ku, Kyoto 606-8502, Japan 
       \and
     Academia Sinica Institute of Astronomy and Astrophysics, 11F of Astronomy-Mathematics Building, AS/NTU, No.1, Section 4, Roosevelt Road, Taipei 10617, Taiwan 
     \and
Research Center for Space and Cosmic Evolution, Ehime University, 2-Bunkyo-cho, Matsuyama, Ehime 790-8577, Japan 
       \and
       Instituto de Astronomıia sede Ensenada Universidad Nacional Autonoma de Mexico Km 107, Carret. Tij.-Ens., Ensenada, 22060, BC, Mexico 
          }

   \date{Received February 2020 }

  \abstract
   { 
   Dusty high-z galaxies are extreme objects with high star formation rates (SFRs) and luminosities. 
   Characterising the properties of this population and analysing their evolution over cosmic time is key to understanding galaxy evolution in the early Universe. 
   } 
   { We select a sample of high-z dusty star-forming galaxies (DSFGs) and evaluate their position on the main sequence (MS) of star-forming galaxies, 
   the well-known correlation between stellar mass and SFR. We aim to understand the causes of their high star formation and quantify the percentage of DSFGs that lie above the MS.  
   }
   { We adopted a multi-wavelength approach with data from optical to submillimetre wavelengths from surveys at the North Ecliptic Pole (NEP) 
 to study a submillimetre sample of high-redshift galaxies. 
 Two submillimetre selection methods were used, including: sources selected at 850$\mathrm{\, \mu m}$ with the Sub-millimetre Common-User Bolometer Array 2) 
 SCUBA-2 instrument and {\it Herschel}-Spectral and Photometric Imaging Receiver (SPIRE) selected sources 
 (colour-colour diagrams and 500$\mathrm{\, \mu m}$ risers), finding that 185 have good multi-wavelength coverage. 
 The resulting sample of 185 high-z candidates was further studied by spectral energy distribution (SED) fitting with the CIGALE fitting code. We derived photometric redshifts, 
 stellar masses, SFRs, and additional physical parameters, such as the infrared luminosity and active galactic nuclei (AGN) contribution.
 }
  { We find that the {\it Herschel}-SPIRE selected DSFGs generally have higher redshifts ($\mathrm{z = 2.57^{+0.08}_{-0.09}}$) than sources that are selected solely by the SCUBA-2 
  method ($\mathrm{z = 1.45^{+0.21}_{-0.06}}$). 
  We find moderate SFRs ($\mathrm{797^{+108}_{-50} M_{\odot}/yr }$), which are typically lower than those found in other studies. 
  We find that the different results in the literature are, only in part, due to selection effects, as even in the most extreme cases, 
  SFRs are still lower than a few thousand solar masses per year. 
  The  difference in measured SFRs affects the position of DSFGs on the MS of galaxies; most of the DSFGs lie on the MS (60\%). 
  Finally, we find that the star formation efficiency (SFE) depends on the epoch and intensity of the star formation burst in the galaxy; the later the burst, 
  the more intense the star formation. 
  We discuss whether the higher SFEs in DSFGs could be due to mergers. 
  }
    {}
   \maketitle 


\section{Introduction}

Submillimetre galaxies (SMGs) are amongst the most luminous dusty galaxies in the Universe \citep{Wilkinson2017}.  
The discovery of this population of bright sources at submillimetre wavelengths (flux densities $\mathrm {F_{850 \mu m} > 2-5 \ mJy}$), 
which is mostly made up of high-z galaxies ($\mathrm{z > 1}$), posed critical questions about the evolution 
of galaxies in the early Universe \citep[e.g.][]{Barger1998, Smail1997, Hughes1998}. 

The launch of the {\it Herschel} Space Observatory \citep{Pilbratt2010} and the observations with the Spectral and Photometric Imaging REceiver instrument
(SPIRE; \citealt{Griffin2010}) have enabled surveys over hundreds of square degrees, allowing searches for rare, exotic objects. 
Galaxies detected by {\it Herschel} tend to be dust rich and to show unusually high stellar masses and star formation rates (SFRs).
For this reason, they are usually classified as dusty star-forming galaxies \citep[DSFG, see][for a review]{Casey2014}. 

The definition of SMGs encompasses cosmological sources selected across the submillimetre wavelength range ($\mathrm{250-1000 \ \mu m}$) \citep{Geach2017}  
and is a subset of DSFGs characterised through a broad range of physical properties \citep{Casey2014}. 
Results have included the first DSFG at z > 6 with an unexpected star formation rate of thousands of solar masses per year \citep{Riechers2013}. 
These types of discoveries have established the existence of large amounts of dust in the early Universe, closer to the epoch of re-ionisation, 
and set essential constraints on theories of galaxy evolution and the cosmic star formation history \citep{Casey2014}. 
It is unclear whether the star formation in DSFGs has been driven by similar physical processes throughout the last 12.8 Gyrs or whether the mechanisms regulating 
the star formation changed or evolved over this time. Some studies have suggested that the SFRs during this time may not have changed significantly and always have had moderate star formation rates \citep{Zavala2018}. 
Therefore, it is important to investigate whether starburst galaxies are typical in the high-z Universe, and if so, whether this implies that major mergers dominated at early times.

Studies of SMGs have used different approaches; some studies emphasise accurate spectroscopic data of a few, sometimes individual, 
sources such as \citet{Riechers2013} or \citet{Riechers2017}. Other studies rely on the selection of hundreds of galaxies via photometric redshift in wide survey fields \citep{Asboth2016}. 
Many of these approaches use only FIR and submillimetre data, which have recently been shown to overestimate photometric redshifts \citep{Ma2019}. 
To correctly quantify the stellar emission without any proxy and to accurately constrain the full spectral energy distribution (hereafter: SED), 
it is essential to combine multi-wavelength measurements ranging from the ultraviolet (UV) to the submillimetre \citep{Buat2014}. 

The main sequence of star-forming galaxies (hereafter: MS) is loosely defined as a correlation between the star formation 
rate and stellar mass ($\mathrm{SFR \propto M_{*} ^{\alpha} }$, \citealt{Noeske2007}). 
It has been demonstrated, by using studies from z=0 to z=6, that the MS is time-dependent \citep{Speagle2014}. 
While there is a consensus on the idea that galaxies lying on the MS form stars over long time-scales ($1-2$ $Gyr$), there are galaxies located above the 
MS with much higher associated SFR. 
These galaxies are often referred to as starbursts and their gas consumption is faster ($0.01-0.1$ $Gyr$) than those star-forming galaxies on the MS \citep{Elbaz2007}. 
There is disagreement regarding the location of SMGs on the MS: while some studies suggest that they are characterised by particularly high levels of SFR, 
lying above the MS of galaxies \citep{Miettinen2017a, daCunha2015}, other studies have found more moderate SFRs following a normal mode of star formation 
\citep[e.g.][]{Koprowski2016, Dunlop2017}. 
A possible explanation for these apparent inconsistencies is suggested by the peculiar morphology of these sources, indicating that mergers play a major role in the 
regulation of  the star-formation \citep{Elbaz2018}, 
enhancing the star formation during these phases \citep{Riechers2017}. 
It is not clear if the nature of extreme galaxies such as HFLS3 ($\mathrm{z=6.3}$ with $\mathrm{SFR \sim 2,900 \, M_{\odot} yr^{-1}}$, \citealt{Riechers2013}) is representative of SMGs
and their reasons for their high SFRs. There is also evidence that starburst galaxies are more compact than MS galaxies and exhibit a polycyclic aromatic hydrocarbon (PAH) deficit \citep{Elbaz2011}. 
Summarising, the mode of star formation in SMGs is still unclear; determining their position on (or off) the MS and the physical reasons behind their mode(s) 
of star formation is key to a better understanding of their nature. 

In this paper we present a study of the high redshift dusty population in a field around the North Ecliptic Pole (NEP) 
having good multi-wavelength coverage. Thanks to the high visibility and relatively low cirrus emission, the NEP represents one of the most important fields for studies at submillimetre wavelengths. 
We selected the high redshift galaxies using surveys at submillimetre wavelengths from {\it Herschel}-SPIRE at 250, 350, 500$\mathrm{\, \mu m}$  
and the Sub-millimetre Common-User Bolometer Array 2 (SCUBA-2) instrument on the James Clark Maxwell Telescope (JCMT) at $\mathrm{850 \, \mu m}$ \citep{Holland2003} combined with multi-wavelength data from ultraviolet, optical and near-mid-far infrared wavelengths.

In Section \ref{observations}, we present and describe the multi-wavelength data sets available in the field.   
In Section \ref{highzcandidates} we present the two methods used to search for and identify possible high redshift candidates which will allow us to compare 'classical' SMGs selected at $\mathrm{850 \, \mu m}$ microns with a population of dusty galaxies selected by their  SPIRE colours. 
The resulting high-z cross-matched catalogue is presented in the same Section \ref{highzcandidates} while in Section \ref{Methodology} the SED fitting methodology is explained. 
In Section \ref{Results} the results are presented including the physical properties of the high-z sample concerning their location in the MS. 
The results are discussed in Section \ref{Discussion} and the conclusions of this work are summarised in Section \ref{conclusions}.    

Throughout this paper, J2000 coordinates and AB magnitudes are used. 
A standard concordance cosmology with $\mathrm{H_{0}=70 \ km s ^{-1} Mpc^{-1}}$, $\mathrm{ \Omega_{m}=0.3}$ and $\mathrm{\Omega_{\Lambda}=0.7}$ 
and a \citealt{Chabrier2003} initial mass function (IMF) is assumed.

\section{Observations}
\label{observations} 

\subsection{The NEP Field}
\label{NEP_field}

\begin{figure}[h!] 
 \centering
 \includegraphics[width=\columnwidth]{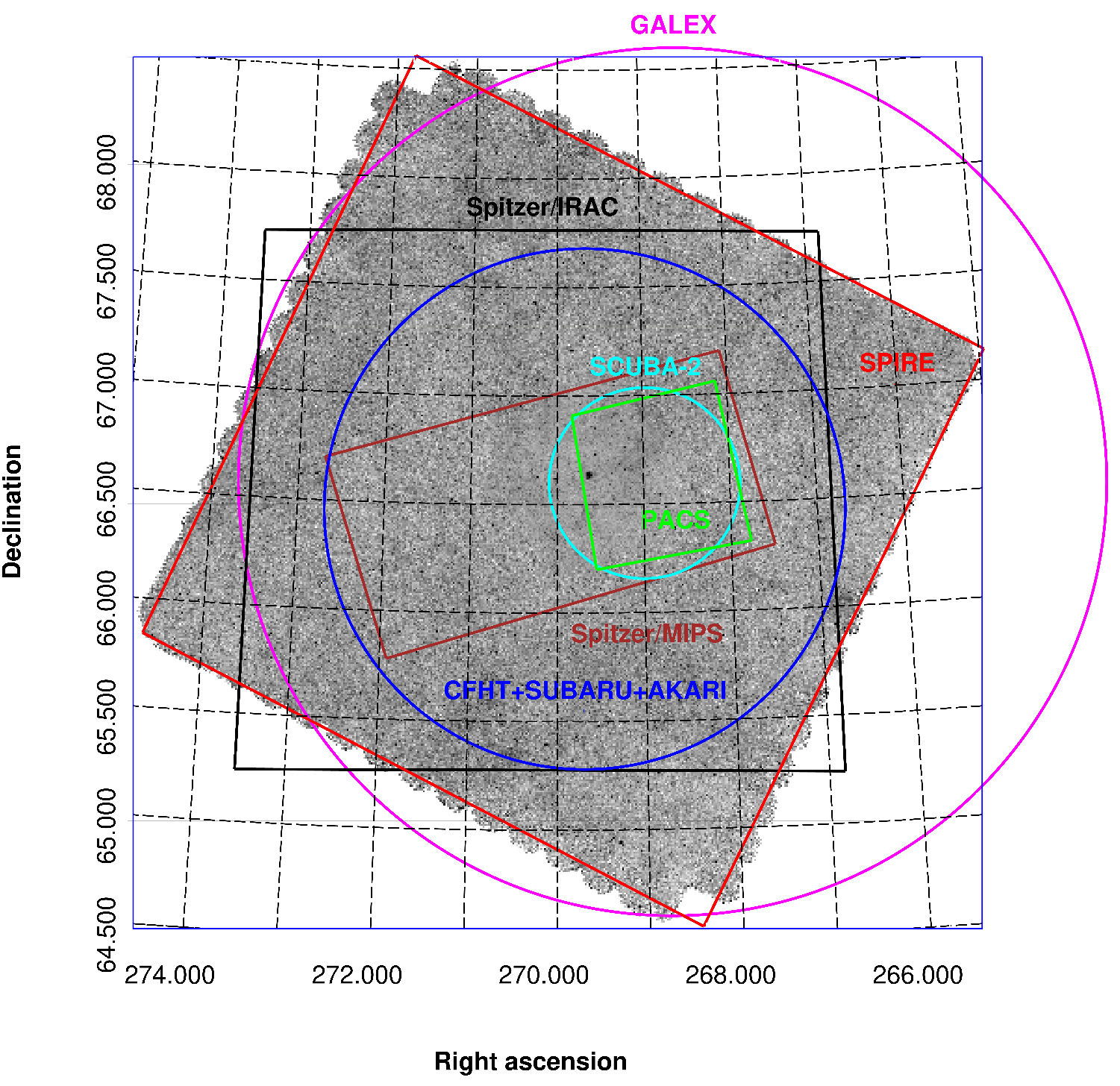} 
 \caption{ Primary surveys located over the NEP field are used in this study. 
 In the inner regions, the cyan circle and the green rectangle represent the SCUBA-2 \citep{Geach2017} and the PACS \citep{Pearson2019} observations respectively.
 The inner dark red rectangle shows the region of the Spitzer/MIPS data, which has no homogeneous borders. 
 The big blue circle represents the area covered by AKARI, CFHT (\citep{Kim2012, Murata2013} and SUBARU (Oi et al. in prep.)) observations.
 The red square corresponds to the area covered by the SPIRE (Herschel)  observations (Pearson et al. 2017, Pearson et al. in preparation).  
 The big magenta circle shows the coverage of GALEX data \citep{Bianchi2014}. }
 
  \label{NEPds9}
\end{figure}

Due to its high visibility, the North Ecliptic Pole (NEP) is a natural cosmological deep field and has been observed by many space telescopes and ground based 
facilities over a wide wavelength range  from radio \citep{White2010} to X-ray wavelengths \citep{Krumpe2015} and is summarised in Table \ref{NEPdatatable} and Figure \ref{NEPds9}. 
The field is included in the multi-wavelength catalogue from the Herschel Extragalactic Legacy project \citep{Shirley2019} 
and has also been selected as a site for the future {\it Euclid} Deep Field \citep{Serjeant2012}.
This legacy field is comprised of two distinct areas defined as the NEP-Deep field, centred at 
$\mathrm{RA=17h \ 55m \ 24s}$ $\mathrm{Dec=66^{\circ} \ 37' \ 32''}$ ($\mathrm{0.54 \ deg^2}$, circular shape) 
and the NEP-Wide field, centred at  $\mathrm{RA=18h \ 00m \ 00s}$ $\mathrm{Dec=66^{\circ} \ 36' \ 00''}$ ($\mathrm{5.4 \ deg^2}$, circular shape) \citep{Matsuhara2006}.

This work uses the most recent data from ultraviolet (UV) to submillimetre wavelengths, primarily concentrating on the submillimetre observations 
from {\it Herschel} and SCUBA-2 described in Section~\ref{observations_IRsubmm} to create a multi-wavelength catalogue of high-z candidates 
(described in Section~\ref{highzcandidates}). We summarise the data used in this paper in the following sections.
For all the catalogues described below, we always considered either total fluxes or magnitudes. 
Furthermore, where available,  the consistency of the fluxes measured with different instruments (or different catalogues) in overlapping filters is confirmed, 
finding good agreement in all cases. This confirms the validity of the cross matches performed. In total, we considered 38 photometric bands with 30 of 
them being non-overlapping bands  (see Table \ref{observations} for details).

\begin{table*}
\begin{center}
 \centering
\begin{tabular}[h!]{|c|c|c|c|c|}
\hline
\textbf{Telescope } & \textbf{Wavelength}  & \textbf{ Area}  & \textbf{Beam size/PSF } & \textbf{Depth}  \\ 
\textbf{[Catalogue] } &   & \textbf{ $\mathrm{[deg^2}]$ }  & \textbf{ [arcseconds]} &  \\ 
\hline
JCMT-SCUBA-2  & 850 $\mathrm{[\mu m]}$ & 0.6 & 15 & 1.2 mJy/beam \\ 
\citep{Geach2017} & & & & ($\mathrm{1 \sigma}$) \\
\hline
{\it Herschel}-SPIRE  & 250 - 350 - 500 $\mathrm{[\mu m]}$  & 9 & 18 - 25 - 36 & 9.0 - 7.5 - 10.8 [mJy] \\ 
(Pearson 2017 / Pearson in prep) & & &  & ($\mathrm{3 \sigma}$) \\ 
\hline
{\it Herschel}-PACS & 100 - 160 [$\mathrm{\mu m}$] & 0.69 & 7-12 & 4 - 8 mJy \\ 
  \citep{Pearson2019} & & & & ($\mathrm{3 \sigma}$) \\ 
\hline
{\it Spitzer}-MIPS & 24 $\mathrm{[\mu m]}$ & 2 & 6 & 30 $\mathrm{\mu Jy}$  \\ 
 \citep{Shirley2019} & & & & ($\mathrm{3 \sigma}$) \\ 
 \hline
 AKARI-IRC  & 2.4 - 3.2 - 4.1 - 7.0 - 9.0 -  & 0.69 & 2.9-6.7 & 11 - 9 - 10 - 30 - 34  \\
\citep{Murata2013}  & 11 - 15 - 18 - 24 [$\mathrm{\mu m}$] &   &  &  57 - 87 - 93 - 256 $\mathrm{\mu Jy}$ ($\mathrm{5 \sigma}$)  \\ 
\hline
AKARI-IRC  & 2.4 - 3.2 - 4.1 - 7.0 - 9.0 - & 5.4 & 2.9-6.7 & 15 - 13 - 13 - 57 - 69  \\ 
\citep{Kim2012}   & 11 - 15 - 18 - 24 $\mathrm{[\mu m]}$  & &   & 100 - 144 - 144 - 275 $\mathrm{\mu Jy}$ ($\mathrm{5 \sigma}$) \\   
\hline  
{\it Spitzer}-IRAC  & 3.6 - 4.5 $\mathrm{[\mu m]}$ & 7.04 & 1.78 & 1.9 - 0.79 $\mathrm{\mu Jy}$ \\ 
 \citep{Nayyeri2018} & & & & ($\mathrm{1 \sigma}$)\\ 
 \hline
 WIRCAM/CFHT  & Y - J - Ks & 0.69 &   & 24 - 23 - 26 mag [AB] \\ 
SuprimeCam/SUBARU & B, V, R, i, z & 0.3 & 0.6 - 1.2 & $\mathrm{\sim 25}$ mag [AB]($\mathrm{5 \sigma}$)  \\
MegaCam/CFHT  & u - g - r - i - z & 0.69 & 0.84 - 0.96 & 26-26-26-25-24 mag [AB] ($\mathrm{5 \sigma}$) \\
\citep{Murata2013} & & & & \\
  \hline
CHFT  & u - g - r - i - z & 2  & 0.84-0.96 & 26-26-26-25-24  \\ 
\citep{Kim2012} & & &                                                    & mag [AB] ($\mathrm{5 \sigma}$) \\
\hline 
SUBARU/HSC &  g - r - i - z - Y  & 5.4  & 0.6 - 1.2  & 27.18 - 26.71 - 26.10 -25.26 - 24.78  \\ 
Oi et al. in prep. & & &                                                                                               & mag [AB] ($\mathrm{5 \sigma}$) \\
\hline
GALEX  & NUV-FUV & All sky  & 4.5 - 6  & $\mathrm{\sim 23}$ \\ 
\citep{Bianchi2014} & & &   & mag [AB] ($\mathrm{5 \sigma}$) \\ 
\hline
     \end{tabular}
       \caption{Overview of the observational data covering the NEP used in this work approximately ordered from long to short wavelengths. The first column indicates the name of the telescope 
       and the reference corresponding to the catalogue used. The second column indicates the photometric bands considered and the third column shows the area covered. 
       The fourth column corresponds to the beam size of the telescopes, and the depth is shown in the fifth column. All of these catalogues have astrometric errors smaller than 2 arcseconds. 
       }
\label{NEPdatatable}      
\end{center}
\end{table*}

%
\subsection{Submillimetre data}
\label{observations_IRsubmm}

The {\it Herschel} Space Observatory \citep{Pilbratt2010} observed the entire NEP region as part of the {\it Herschel} Open Time Program, with the SPIRE instrument (OT2sserje012; P.I Serjeant). 
The SPIRE observations cover an area ($\mathrm{\sim 9 \, deg^{2}}$) at 250$\mathrm{\, \mu m}$, 350$\mathrm{\, \mu m}$ and 
500$\mathrm{\, \mu m}$. The SPIRE catalogue is described in \citet{Pearson2017}, Pearson et al. in prep., and contains 4820 sources, 
extracted via an iterative  source extraction procedure (see Section~\ref{cross-match}) to optimise the detection of individual sources at $\mathrm{3\sigma}$ 
higher than the extragalactic confusion noise limits of 5.8, 6.3 and 6.8 mJy/beam ($\mathrm{1\sigma}$)  at 250, 350  and 500$\mathrm{\, \mu m}$ respectively \citep{Nguyen2010}. 

Longer wavelength submillimetre data is also available through the SCUBA-2 cosmology legacy survey \citep{Geach2017} at 850$\mathrm{\, \mu m}$ using the Sub-millimetre 
Common-User Bolometer Array 2 (SCUBA-2) instrument, operating at the James Clerk Maxwell Telescope (JCMT). This survey covered the NEP-Deep area (see Figure \ref{NEPds9}) area reaching a sensitivity of 
$\mathrm{2 mJy beam^{-1}}$ \citep{Geach2017}.    
The SCUBA-2 catalogue at the NEP contains 330 sources which have at least a 3.5$\mathrm{\sigma}$ detection. 
In this study, we considered SCUBA-2 de-boosted fluxes, which allow for a proper subtraction of the background and make possible to obtain more precise estimates of the total fluxes 
(as for the other catalogues used in this paper). These two data sets will form the basis of our selection methods that will be used to select high-z sources over the NEP area.

%
\subsection{Far infrared data}
\label{observations_farIR} 
The NEP-Deep region was also observed in the same {\it Herschel} Open Time programme  with the  Photodetector Array Camera and Spectrometer 
(PACS) instrument \citep{Poglitsch2010} at 100$\mathrm{\, \mu m}$ and 160$\mathrm{\, \mu m}$. The survey area covered approximately ($\mathrm{\sim 0.69 \ deg^{2}}$) 
closely overlapping the region of the sky covered with SCUBA-2 (see Figure \ref{NEPds9}). The PACS source extraction and catalogue is presented in \citet{Pearson2019} 
and contains 1385 and 630 sources to limiting fluxes of  4 and 8 mJy at 100$\mathrm{\, \mu m}$ and 160$\mathrm{\, \mu m}$ respectively (see Table \ref{NEPdatatable}).  
The addition of the PACS data into our data set is important to better constrain the far-infrared (FIR) spectral peak. 

%
\subsection{Mid- and near-infrared data} 
\label{observations_MIRNIR} 

The NEP field has been extensively observed by the {\it AKARI} satellite \citep{Murakami2007} as an {\it AKARI} legacy field \citep{Matsuhara2006}. 
The NEP was in the continuous viewing zone of  {\it AKARI} and was visible several times per day allowing deep near-to mid-IR imaging of the region. 
The NEP was observed in all nine bands of the {\it AKARI} IRC instrument \citep{Onaka2007}:  N2 ($\mathrm{2.4 \mu m}$), N3 ($\mathrm{3.2 \mu m}$),  N4 ($\mathrm{4.1 \mu m}$), S7 ($\mathrm{7.0 \mu m}$), 
S9W ($\mathrm{9.0 \mu m}$), S11 ($\mathrm{11 \mu m}$), L15 ($\mathrm{15.0 \mu m}$), L18W ($\mathrm{18.0 \mu m}$), L24 ($\mathrm{24 \mu m}$). 
Dedicated surveys were made over both the NEP-Deep and NEP-WIDE regions. The NEP-Deep catalogue of \citet{Murata2013} includes 11,349 sources over 0.5 $\mathrm{deg^{2}}$ 
with at least a detection in one of the AKARI bands. The use of a high S/N cut (5$\mathrm{\sigma}$) allows for a low false detection rate (<0.3 \%). 
The NEP-Wide catalogue \citep{Kim2012} contains 114,974 infrared sources over  a larger area of ($\mathrm{5.4 \ deg^{2}}$) but to a shallower depth than that of the \citet{Murata2013} 
NEP-Deep catalogue (see Table \ref{NEPdatatable} for comparison). 

{\it Spitzer}-MIPS data at 24$\mathrm{\, \mu m}$ over a limited area of the NEP region is available through the {\it Herschel} Extragalactic Legacy Project (HELP)\footnote[1]{http://herschel.sussex.ac.uk/}, 
an ongoing program aiming to produce a multi-wavelength database sets for all  Herschel extragalactic surveys \citep{Shirley2019}. HELP uses the XID+ \citep{Hurley2017} 
algorithm, a deblending prior-based source extraction tool to perform photometry at the positions of known sources in far-infrared confusion dominated maps from shorter wavelength 
(MIPS 24 $\mathrm{\, \mu m}$).  The algorithm  based on a probabilistic Bayesian framework which allows for the inclusion of prior information (beyond source positions and fluxes) 
to obtain the full posterior probability distribution on flux estimates rather than just the maximum likelihood. The availability of data with a higher spatial resolution at shorter wavelengths 
(optical, radio and mid-near-infra-red) enables the de-blending of Herschel maps. We note that, although SPIRE data is also included in the HELP catalogue, 
the photometric errors in the SPIRE band are large (the relative error is near 40\% on average in the SPIRE bands). Moreover, the NEP region the MIPS observations cover about 
$\mathrm{\sim 2 deg^{2}}$ for a total of 20 mosaics, which represent approximately a quarter of the area covered by SPIRE in the NEP field. However, the HELP catalogue is extremely valuable and we exploit this catalogue to check the reliability of the techniques that we used to cross-match all the multi-wavelength 
observations (see Section \ref{cross-match}).

Observations of the NEP region at near-infrared wavelengths were also made by {\it Spitzer} during its warm mission phase with the IRAC instrument. The survey of \citet{Nayyeri2018} 
cover an area of a total area of $\mathrm{7.04 \, deg^2}$, at 3.6$\mathrm{\, \mu m}$ (IRAC1) and 4.5$\mathrm{\, \mu m}$ (IRAC2).  The final catalogue contains 380,858 sources \citep{Nayyeri2018}. 
The band-pass of the IRAC1 and IRAC2 filters is similar to that of the AKARI N3 and N4 filters respectively. However, the observations in the \textit{Spitzer}-IRAC bands are deeper (1.29 and 0.79 $\mathrm{\mu Jy }$ than the corresponding AKARI bands ($\mathrm{\sim 13 \ \mu Jy }$ for N3 and N4). The deeper IRAC observations allow us to 
obtain additional information avoiding mismatches in our high-z multi-wavelength catalogue (see Section \ref{Results}).  
The IRAC catalogue of \citet{Nayyeri2018} includes aperture photometry considering three different apertures: automatic aperture ('AUTO'), 4 arcseconds ('Aperture 1'), 
6 arcseconds ('Aperture 2'). The fluxes computed inside these three apertures were compared to those reported for AKARI N3 and N4 in the catalogues \citep{Murata2013, Kim2012}. 
As a result of this comparison, we obtained the best agreement when considering the automatic apertures in \citealt{Nayyeri2018}, while an offset is measured for all the fixed apertures. 
For these reasons in this study, only the automatic aperture was considered (see our high-z DSFGs cross-matched catalogue Section \ref{cross-match}).

%
\subsection{Optical data} 
\label{observations_optical}

Optical follow-up surveys have been made over both the NEP-Deep and NEP-Wide survey areas. The {\it AKARI} catalogue of \citealt{Murata2013} 
accurate cross-matching of the IR detections with their optical counterparts using B, V, R, i, z, NB711 from SUBARU-SuprimeCam,   u, g, r, i, z from MegaCam-CFHT and Y, J, Ks from WIRCAM-CFHT 
data (see \citealt{Murata2013} for details). The NEP-Wide catalogue of \citet{Kim2012}, besides of the IR data, includes over $\mathrm{\sim}$ 110,000 crossed-matched 
sources with u, g, r, i, and z band data from MegaCam-CFHT observations to a detection limit down to  26 magnitudes (4$\mathrm{\sigma}$) 
covering $\mathrm{\sim 2 \ deg^{2}}$ of the central part of the NEP.  

Additional optical data from Hyper Suprime camera (HSC) on the SUBARU telescope(Oi et al. in preparation) were also included in our multi-wavelength dataset. 
This data cover $\mathrm{\sim 6 \ deg^2}$ 
in the g, r, i, z, Y optical and near-IR bands. The final catalogue contains over two million sources with a 5$\mathrm{\sigma}$ limiting magnitudes of 
27.18, 26.71, 26.10, 25.26, and 24.78 mag [AB] in the five bands respectively \citep{Goto2019}. The catalogue includes 89,178 {\it AKARI}-IRC infrared sources that are already cross-matched with the observations in the optical bands. 

In total, there is a broad coverage across the optical and near-mid-IR part of the spectrum, covering 28 bands in the NEP-Deep field and 14 
in the NEP-Wide field. The main characteristics of these catalogues are summarised in Table \ref{NEPdatatable}.

%
\subsection{Ultraviolet data}
\label{UVdata}

In the UV spectral range, the NEP field is covered by the Galaxy Evolution Explorer (GALEX) space telescope \citep{Martin2005}, at both far-ultraviolet (FUV) and near-ultraviolet (NUV),  
and contains 36,263 sources as a part of the all-sky survey described in \citealt{Bianchi2014}. The observations are relatively shallow with a depth of 23 magnitude; 
consequently, a relatively low number of matches are expected, since most of our sources would be too faint to be detected by GALEX (see Section \ref{cross-match} for more details).

%
\subsection{Spectroscopic redshift catalogues}
\label{specredshift}

Very limited spectroscopic data is available in the NEP region, mostly  located in the NEP-Deep survey area. 
Some of this data were collected in the HELP catalogue of  \citet{Shirley2019} and is summarised in \citet{Shim2013}. 
The quality of the redshifts in the HELP catalogues is defined by a flag assuming a value ranging from 1 (unreliable) to 5 (high quality). 
There are  $\mathrm{\sim 1100}$ sources with spectroscopic redshifts in the NEP region, however, the quality of these spectroscopic redshifts are all flagged as $\mathrm{\leq}$3.

In addition to the HELP spectroscopic sub-sample, we considered the catalogue of Kim et al. (in preparation). 
However, the majority (84 \%) of the sources reported in this catalogue are located at $\mathrm{z < 1}$. 

We concluded that the number of sources with an available reliable spectroscopic measurement is not sufficient to complete a robust statistical analysis. 
For this reason, in this study we relied on the photometric redshifts (see Section \ref{photometricredshiftresults}).

\section{Selection of high-z candidates and multi-wavelength catalogue}
\label{highzcandidates}

%
\subsection{Selection of high-z candidates} 
\label{selection}

In this work, the analysis focuses on two samples of high-z galaxies selected using different methods. 
In the first sample, high-z candidates were selected using SPIRE colour criteria, while in another case, the selection was based on SCUBA-2 fluxes. 

\begin{center} 
\begin{figure} 
\includegraphics[width=1.0\linewidth]{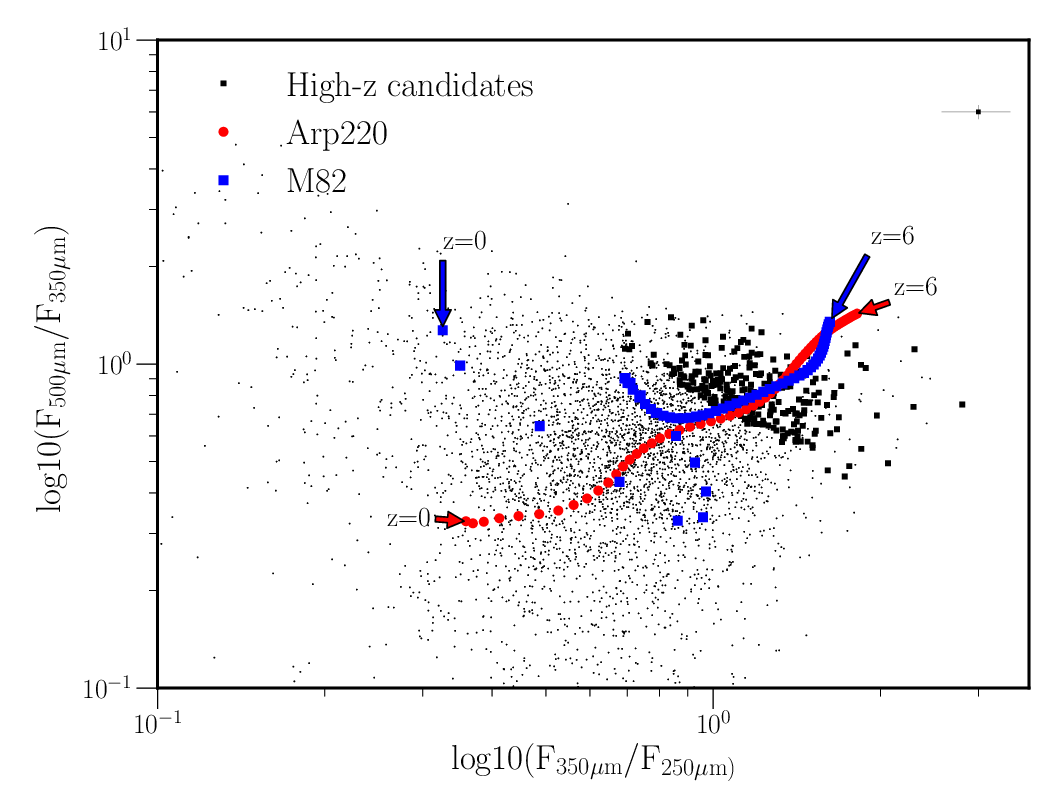} 
  \caption{  Colour-colour diagram of the SPIRE fluxes with the tracks of the local starburst galaxies M82 (blue) and Arp220 (red) over-plotted.
   The tracks represent the observed colours as expected from the same SEDs observed at different redshifts from $\mathrm{z=0}$ to $\mathrm{z=6}$ 
   with redshift steps of $\mathrm{\Delta z = 0.1}$ 
   The bold points are the high-z candidates which lie in the same colour space than the evolved tracks at z > 2. 
   The upper right area of the plot shows the average error bar for the high-z candidates. 
   all these methods are based on MBB models,
where the redshift evolution of the SEDs is not included. Therefore, more sophisticated SED templates should be used to investigate the correlation between redshift and colours. }
\label{cc_diagram}
  \end{figure}
  \end{center}

The peak of the far-IR emission due to dust, approximately located at rest-frame $\mathrm{\lambda \sim 100 \mu m}$, can be used as a rough redshift indicator.
To this purpose, observed {\it Herschel} colours are used to select potential high-z sources as these colours allow sampling of the peak of the dust emission at different redshifts. 
Submillimetre-FIR colour-colour diagrams are commonly used to find high-z candidates \citep[e.g.][]{Amblard2010, Ivison2012}. 
A similar concept is utilised by the 500 $\, \mu$m risers method, where sources exhibiting increasing SPIRE fluxes moving from $\mathrm{250 \, \mu m}$  to $\mathrm{350 \, \mu m}$  
and then onwards to $\mathrm{500  \, \mu m}$  are more likely to be located at very high redshift, since the dust emission peak appears to be redshifted 
to $\mathrm{\lambda > 500 \mu m}$ ($\mathrm{z >4}$). 
The  500$\mathrm{\, \mu m}$ riser criterion is successful in selecting massive high-z sources \citep{Donevski2018}. 
The use of submillimetre colours to select high-z candidates relies on the redshifting of the source SED through the observation bands, defining the colour-colour parameter space where 
potential high-z sources will lie (see Figure \ref{cc_diagram}). We therefore define a SPIRE colour criteria following \citet{Amblard2010} 
of $\mathrm{F_{350 \, \mu m} > 35 \, mJy}$ and $\mathrm{F_{500 \, \mu m} / F_{250 \, \mu m}}$ > 0.75 
in order to select a total of 268 high-z candidates from our SPIRE catalogue, these sources are expected to lie at at $\mathrm{z > 2}$. 
In addition, the $\mathrm{500 \, \mu m}$ riser colour ($\mathrm{F_{500 \, \mu m} > F_{350\mu m}}$  $\mathrm{> F_{250\mu m} }$) criterion was applied to select  41 high-z candidates, 
23 of which not already included in the high-z sample. 

The previous criteria are based on modified black body models. We added another criterion where the redshift evolution of the SED taken intro account \citep{Yuan2015}. 
We find 102 new sources that follow the colours $\mathrm{F_{500 \, \mu m}/F_{350 \, \mu m} > 1.1 \cdot (F_{250 \, \mu m}/ F_{350 \, \mu m})- 0.15}$ \citep{Yuan2015}. 
This use of several colour criteria allow us to obtain a high-z sample, not only with extreme sources at z>4 (e.g. \citet{Asboth2016}), 
to study the properties of DSFGs and their evolution over redshift (see Section \ref{Results})).

Our second high-z sample is constructed independently by applying a threshold to the SCUBA-2 fluxes. 
In particular, we selected only sources characterised by $\mathrm{F_{850  \, \mu m} > 3 mJy }$, 
from the catalogue of \citet{Geach2017}.  
In total, 422 sources comply with this criterion which is most likely to select sources at $\mathrm z \sim 2$ and less massive than 
those selected using the SPIRE colours \citep{Casey2014}. 

Using these two criteria; SPIRE colours and SCUBA-2 flux, we selected a total of 291 high-z candidates using SPIRE colour criteria 
(268 with Amblard criteria, 23 risers and 140 with Yuan criteria) and 422 sources using the threshold applied to SCUBA-2 fluxes. 
We refer them as SPIRE selected sources, and SCUBA-2 selected sources, respectively. 
We noticed and expected that the number of potential high-z sources is significantly reduced when we cross-matched the submillimetre detections with the rest of the multi-wavelength data 
after checking the fitted SEDs for possible mismatches (see Sections \ref{cross-match} and \ref{Methodology}, respectively).

%
\subsection{Multi-wavelength high-z catalogue} 
\label{cross-match}

The high-z candidates selected were cross-matched with the data available in the NEP region (see Section \ref{observations}) 
associating every detection in each band to the closest counterpart in the others.  
The SPIRE maps suffer from lower resolution and photometric confusion leading to the blending of multiples sources due to 
the relatively large beam areas of {\it Herschel}-SPIRE, ranging from 17.6 - 35.2 arcseconds from 250 - 500 $\, \mu$m  respectively \citep{Griffin2013}. 
To minimise these effects and to aid in our multi-wavelength cross-matching, the SPIRE catalogue of Pearson et al. (in preparation) was created 
via an iterative  extraction procedure to optimise the detection of individual sources.  A similar approach has been used to successfully extract sources for the {\it Herschel} 
HeViCS programme \citep{Davies2010, Pappalardo2015}, identifying bright peaks in the maps and then fitting with the SPIRE 250 $\, \mu$m PSF. 
In an iterative process, the most luminous sources extracted are subtracted from the original image, then, the source extraction is performed on this new cleaned image, 
but this time considering a lower detection threshold. 
The iteration is repeated until the extraction threshold reaches the level of the noise. Photometry at 350 and 500 $\, \mu$m is obtained considering the 250 $\, \mu$m prior positions.  
Similar approaches have been adopted in other other large-area surveys with {\it Herschel}  \citep[e.g.][]{Clements2010, Rigby2011}. 

As stated above, the catalogues are matched by pairing the closest counterparts in different photometric bands. To this purpose, when matching two bands, we considered a search radius 
corresponding to $\mathrm{0.5*FWHM}$ of the band with the larger PSF. A similar technique is successfully adopted in \citet{Baronchelli2016, Baronchelli2018}.

%
\subsubsection{SPIRE colour selected sample} 

The multi-wavelength catalogue for the high-z DSFGs selected via SPIRE colours was made using the 250$ \, \mu$m SPIRE position (the highest resolution band) 
and their closest longest wavelength band counterparts. Counterparts in the SCUBA-2 catalogue of \citet{Geach2017} were cross-matched with the 250$\mathrm{\, \mu m}$ band 
using a search radius smaller than nine arc-seconds, 
(half of the SPIRE  beam FWHM at 250$\mathrm{\, \mu m}$). Most of the SPIRE selected high-z candidates (74\%) 
have an 850$\mathrm{\, \mu m}$ counterpart following the classical definition of a SMG ($\mathrm{F_{850} > 3 \, \mu mJy}$). 

PACS data was incorporated by producing the photometry in the PACS maps at the SPIRE 250$\mathrm{\, \mu m}$ position of the high-z candidate. 
The photometry was produced using an identical procedure to that of \citealt{Pearson2019} in both PACS maps at 100$\mathrm{\, \mu m}$ and 160$\mathrm{\, \mu m}$. 
In this way, a PACS flux (or at least an upper limit) can be recovered for all our high-z candidates. As a means of validation of our PACS photometry, 
we compared the brighter PACS fluxes to confirm that they have the same measured flux as by cross-matching SPIRE candidates directly with the PACS catalogue of \citealt{Pearson2019}. 
The addition of PACS photometry produces complete coverage of the FIR peak for the sources located in the deep field. 

Infrared counterparts in the NEP-Deep \citet{Murata2013} and NEP-Wide \citet{Kim2012} catalogues (including their optical counterparts in the catalogues) 
were cross-matched with the high-z catalogue using a search radius is 9 arcseconds (half of the SPIRE $\mathrm{250 \mu m}$ beam) 
considering the possible counterparts inside this search radius. For both catalogues, in order to select the correct counterpart, 
the longest wavelength mid-infrared $\mathrm{15 \mu m}$ and $\mathrm{18 \mu m}$ bands were used.  

For the optical  Subaru-HSC data, which covers both deep and Wide field, the mid-infrared bands are used effectively as an 'IR-bridge' \citep{Baronchelli2016}, 
choosing the optical detection associated with the brighter IR detection. The two {\it AKARI} catalogues and optical data overlapped, the associations were compared band by band, finding a good
agreement of fluxes between them. 

The {\it Spitzer}-IRAC catalogue with 3.6 and 4.5$\mathrm{\, \mu m}$ bands \citep{Nayyeri2018} was cross-matched using a search radius 
of 9 arcseconds with the 250$\mathrm{\, \mu m}$ SPIRE band, finding 261 identifications in at least one of the IRAC bands. 
We note the same percentage of matches in the deep than in the wide field (80\%) although the SPIRE data is deeper in the NEP-Deep region. 
These sources were analysed with CIGALE in Section \ref{Methodology}.  However, only the high-z candidates with more data than SPIRE and IRAC detection were included in Section \ref{Results}. 
Where sources had flux detections in both the {\it AKARI}-IRAC and {\it AKARI}-IRC overlapping bands, the source fluxes were verified to ensure that the same source was associated in both cases.

As an additional level of validation, the HELP catalogue was also used to check any miss-associations between IRAC and SPIRE data, since the HELP catalogue contains SPIRE sources. 
The HELP catalogue was cross-matched with our high-z catalogue based on the 250$\mathrm{\, \mu m}$ positions using a 
3 arcseconds search radius. We find that the associations between the two SPIRE catalogues agree in flux, within the errors (see Appendix~\ref{appendix_herschel} and Figure \ref{XID_herchel}) 
and we find extremely few sources with double counterparts. This flux agreement not only shows that we are selecting the same sources but also emphasises that the source identification and the extraction method used in Pearson et al. in prep, 
which covers four times more area than in the HELP catalogue (see Table \ref{NEPdatatable}) successfully deblends the SPIRE sources. 
The HELP catalogue also includes Spitzer 24$\mathrm{\, \mu m}$ data and as a result of this cross-match, ten identifications at 24$\mathrm{\, \mu m}$  were incorporated into our multi-wavelength catalogue.

Finally, the UV GALEX catalogue was cross-matched with the optical counterparts with a search radius smaller than 4 arcseconds 
(half of the beam radius). However, only 13 sources of the final sample have a UV counterpart, due to the shallow coverage of GALEX. 

Summarising, the final high-z DSFG catalogue contains 25 high-z candidates with coverage over 38 photometric bands approximately defined over the NEP-Deep area and 266 
high-z candidates with coverage over 26 bands in NEP-WIDE area.  In total, our high-z  DSFG catalogue contains 291 (266+25) sources.

The SPIRE and the {\it AKARI-IRC} catalogues are deeper in the NEP-Deep region and,  
due to the differences in both the number of photometric bands and depth, the analysis in Section \ref{Results} was produced separately for the Deep and the Wide fields.

%
\subsubsection{SCUBA-2 selected sources}

The 288 SCUBA-2 selected galaxies were cross-matched with the rest of the data described in Section \ref{observations} using the original SCUBA-2 positions as the reference. 
The SPIRE catalogue was cross-matched by using a nine arc-second search radius. In total, three double detections between the SPIRE and SCUBA-2 catalogues were found. 
In these cases, the closest counterpart was selected. For the SCUBA-2 sources with no cross-match in the SPIRE catalogue, the photometry was calculated using the SCUBA-2 position in each of the SPIRE maps, 
in the case of no detection, the upper limits for the 3 SPIRE bands were incorporated. 

The PACS data, at 100 and 160$\mathrm{\, \mu m}$, were included by cross-matching with the SCUBA-2 positions using a 7.5 arcsecond search radius finding identifications for 13\% of the sources. 

The SCUBA-2 survey lies within the {\it Spitzer}-MIPS 24$\mathrm{\, \mu m}$  coverage (see Figure \ref{NEPds9}), and all 288 SCUBA-2  sources were cross-matched with 24$\mathrm{\, \mu m}$ 
HELP catalogue using a 7.5 arcsecond search radius (half of the SCUBA-2 beam size). We find 123 SCUBA sources with a reliable detection, the rest (165 high-z candidates) were not included in the analysis since they have either an unreliable  24$\mathrm{\, \mu m}$ flux or a lack of counterpart. 
Using the 24$\mathrm{\, \mu m}$ source position, these data were taken as an 'IR-bridge'  \citep{Baronchelli2016} for incorporating the ancillary data at shorter wavelengths. 
The {\it AKARI}  infrared and associated optical CFHT and SUBARU data (SuprimeCam) from \citet{Murata2013} over the NEP-Deep region were incorporated with a 3 arcsecond search radius. 
The deeper SUBARU-HSC data (Oi et al. in prep.)  were incorporated by using with the same search radius.  
Where the optical bands overlapped in two or more of the \citet{Murata2013} {\it AKARI}, SUBARU-HSC, HELP catalogues the positions and photometry were found to be in good agreement, 
reinforcing our cross-matching confidence.

The final sample of SCUBA-2 sources (SMGs) contains 123 sources, covering 37 photometric bands, similar to the DSFGs selected via the SPIRE colours in the NEP-Deep field. 
Therefore, the results of the two samples can be compared to further decrease possible instances of a miss-match and removing less reliable associations from the final sample (see Section \ref{Results}).

\section{Methodology: SED fitting with CIGALE} 
\label{Methodology}

The physical properties of our high-z sample were calculated using SED fitting techniques with the CIGALE software (Code Investigating GALaxy Emission, 
\citep{Burgarella2005, Noll2009, Boquien2019}\footnote[2]{https://cigale.lam.fr/}). 
CIGALE is a SED fitting code based on the physical properties of galaxies that covers the spectrum from UV to radio wavelengths, 
taking into account the balance between the energy emitted in the UV-optical, absorption by dust, and the reprocessing of the starlight into the IR.  
In this Section, we describe the modules tested within CIGALE. Table \ref{Cigaletable} describes the main parameters of each module that was finally selected. 

\begin{table*}
\begin{center}

\begin{tabular}[h!]{|l|l|l|}
\hline
\textbf{CIGALE module} & \textbf{Main parameters} & \textbf{Description} \\ 
\hline 
SFH    & $\mathrm{f_{burst}= 0.0, 0.05, 0.1, 0.2, 0.3, 0.4, 0.5}$  &  Mass fraction of the late burst population \\ 
two decreasing exponentials         & $\mathrm{age_{burst} = 10, 30, 50, 70, 100, 120, 150}$ [Myr]    &  Age of the late burst   \\
                     & $\mathrm{age = 500, 1000, 2000, }$   &    Age of the main stellar population in the galaxy  \\
                     & 3000,  4000, 5000 [Myr]      & \\
&  $\mathrm{\tau_{main} = 1000, 3000, 5000 }$ [Myr] & e-folding time of the main population   \\                      
& $\mathrm{\tau_{burst} = 3000, 9000 }$ [Myr] & e-folding time of the starburst  \\   
 
 \hline
Single stellar population models & $\mathrm{Z_{\odot}}$ & Metallicity \\                                
\citep{Bruzual2003} & IMF = \citet{Chabrier2003} & Initial mass function \\
\hline
Nebular emission  & $\mathrm{log U = -2}$ & Ionisation parameter   \\
\citep{Inoue2011} &  & \\
\hline
Dust attenuation       &  $\mathrm{Av = 0.3, 2.3, 3.8}$ &  V-band attenuation in the birth clouds \\
\citep{Charlot2000}    &  $\mathrm{slope_{BC} = -0.7}$ & Power law slope in the birth clouds \\
                       &  $\mathrm{slope_{ISM} = -0.7}$ & Power law slope in the  \\
      \hline
Dust emision models     &  $\mathrm{q_{pah} = 1.12, 2.50, 3.9}$   & Mass fraction of PAHs \\ 
\citep{Draine2014}     &   $\mathrm{u_{min} = 5, 10, 25, 40, 50}$        & Minimum radiation field \\
 \hline
AGN component        & fracAGN =  0, 0.05, 0.1, 0.15  & AGN fraction: contribution of the AGN  \\ 
& 0.2, 0.3, 0.4, 0.6 & to the total $\mathrm{L_{IR}}$ \\ 
  \citep{Fritz2006}  &    &    \\
   & $\mathrm{r_{ratio} = 60}$   & Opening angle of the dust torus \\ 
                     &  psy = 0.001, 89.9  & Angle between AGN axis and line of sight \\
& $\mathrm{\tau _{9.7} = 1.0, 6.0}$ $\mathrm{\mu m}$ &  Optical depth at 9.7 \\
                     \hline
      \end{tabular}
       \caption{ The main modules and input parameters used in CIGALE for the analysis of the high-z sample. The first column lists the CIGALE module, 
       the second column shows the main parameters and the range of values selected. The third column provides a brief description of each parameter. }    
\label{Cigaletable}      
\end{center}
\end{table*} 

\begin{figure}
\centering
    \includegraphics[width=1.0\linewidth]{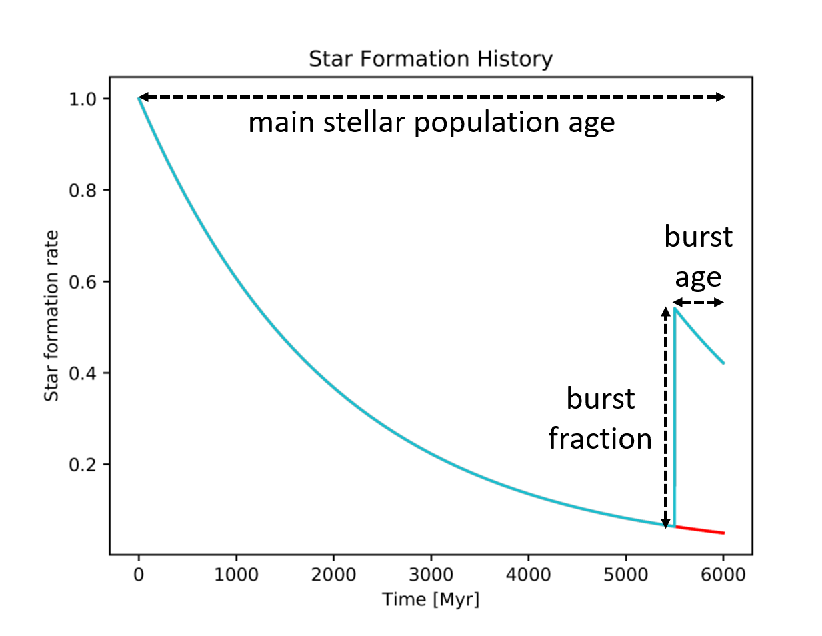} 
\caption{ Star formation history (SFH) adopted in this work: a double-exponential decreasing with starburst age. 
A decreasing exponential describes the star formation with the age of the main stellar population of the galaxy ($\mathrm{t_{age}}$) plus a burst of star formation 
at a specific epoch ($\mathrm{t_{burst}}$)  with a burst fraction ($\mathrm{f_{burst}}$) defining the intensity of the burst.}
\label{SFH_2exp_starburst}
\end{figure}

As the first step, CIGALE builds a stellar component defining the energy emitted by stars. 
This stellar component takes as an input an assumed star formation history (SFH) defined in an independent module in CIGALE, which is crucial in estimating the SFR. 
Several SFHs describing the star formation of galaxies across cosmic time are available considering different physical processes depending
on the nature of the galaxy. DSFGs tend to have high star formation, and the possibility of having a burst of star formation was evaluated 
by using a double decreasing exponential SFH defined as: 

\begin{equation}
 SFR \propto 
 \begin{cases}
 \mathrm{ \exp(-t / \tau_{main})  \hspace{2.95cm} : \  t < t_{main} - t_{burst}} \\
 \mathrm{ \exp(-t / \tau_{main}) + f_{bust} \exp(-t/ \tau_{burst}) \ : \ t \geq t_{main} - t_{burst} }
 \end{cases} 
  \end{equation}
where $\mathrm{t_{burst}}$ is the age of a second episode of star formation in the galaxy and  $\mathrm{t_{main}}$ corresponds to the age of the main population of the galaxy. 
This SFH assumes a decreasing exponential with a presence of a starburst if the fraction of the burst $\mathrm{f_{burst}>0}$ or only a decreasing exponential if there is no starburst 
(see Figure \ref{SFH_2exp_starburst}), where $\mathrm{\tau_{main}}$ and $\mathrm{\tau_{burst}}$ are the e-folding times of the main and burst populations respectively.

The exponential SFH provides the best estimates of Mstar and SFR, but the ages may not be as realistic as with other SFH \citep{Ciesla2015}. 
We applied a broad range of values in this model to allow the possibility of either a moderate or an intense burst of star formation. Besides, the broad range of ages allows different redshifts. 

After computing the SFH, a single stellar population library is needed for modelling the stellar spectrum. 
The widely used \citet{Bruzual2003} model assumes a single stellar population 
by isochrone synthesis,
which takes stars of the same age and integrates their spectra to compute the total flux. The disadvantage is that the isochrones are calculated in discrete steps in time and therefore, 
any stellar evolution more rapid than these time steps is not well represented.
For this reason, we compared the alternative \citealt{Maraston2005} models, which are fuel-consumption based algorithms that follow a 'turn off' from the main sequence of stars. 
In both stellar population models, we assume the initial mass function given by \citet{Chabrier2003} and a solar metallicity $\mathrm{Z_{\odot}}$ (see Table \ref{Cigaletable}) 
as a standard assumption that allow us to compare our results with the rest of the literature. 

The nebular emission module, based on the empirical templates of \citet{Inoue2011}, describes the spectral (line) emission, which is essential in star-forming regions and 
high-z redshift galaxies \citep{Stark2013}. 
This emission has a direct impact on the SED modelling, and it can increase the average flux by up to 10\% in the case of the most energetic spectral lines \citep{Noll2009}. 
We find that SFR is lower when the nebular emission is considered, since a fraction of the radiation contributes as nebular emission instead of the continuum emission of the galaxy.

For the dust attenuation, three different options were investigated. The most common attenuation law for star-forming galaxies is the \citet{Calzetti2000} law with an extension to shorter wavelengths in the spectrum provided by \citet{Leitherer2002}. The second model,  proposes a simpler attenuation power-law \citep{Charlot2000}, that assumes that all the stars are attenuated by diffuse dust in the same manner. However, a single attenuation curve proves to be insufficient to describe DSFGs,  particularly at high-z \citep{Noll2009} and it is not appropriate for our sample. 
The third model described in  \citet{Boquien2019} extends the model of  \citet{Charlot2000} and combines both the birth cloud attenuation and the interstellar medium attenuation each represented by a power law, and therefore makes a distinction between the young and old stellar emission. 
This has been shown to characterise well the emission from ULIRGS at $\mathrm{z \sim 2}$ \citep{LoFaro2017}. 

CIGALE was tested with all three dust attenuation options. Although we did not find any significant differences in the goodness of the fits between these models, the third option was selected on the basis that it makes a distinction between the stellar emission components which is more appropriate for IR galaxies \citep{Buat2018, Malek2018}.
The main parameters that define the model are the attenuation ($\mathrm{A_{V}}$) and the slopes of the 2-power law 
attenuation for the birth cloud and the interstellar medium (see Table \ref{Cigaletable}). 

Models for the emission at mid-infrared (MIR) wavelengths are complex since they require the addition of PAH emission (dust features which are extremely grain size dependent). 
Our sample of dusty high-z galaxies was tested with both the \citet{Draine2014} and \citet{Schreiber2016} models. \citet{Draine2014} takes into account the emission from small dust grains and the characteristics of intense MIR emission 
(extreme heating environments tend to stop the process of dust formation) typically present in star-forming galaxies. 
The model is carefully described in \citet{Draine2007} with and the updated version that allows higher dust temperatures \citep{Draine2014}. 
The main parameters are the mass fraction of the PAH population ($\mathrm{q_{pah}}$) and the minimum radiation field ($\mathrm{U_{min}}$) which is related to the dust temperature \citep{Aniano2012}. We set big $\mathrm{U_{min}}$ values to allow 
higher temperatures (see Table \ref{Cigaletable}). 
The \citealt{Schreiber2016} model is a simpler alternative with regards to the PAH emission, being instead more focused on the FIR-submillimetre peak.  
We used the bayesian interference factor (BIC) to compare the goodness of the fit of two models. The difference between \citealt{Schreiber2016} \citealt{Draine2014} is $\mathrm{\triangle BIC \sim 6}$ in the advantage of \citealt{Draine2014} meaning that is significantly better \citep{Ciesla2018}. 
In addition, the models of \citet{Casey2012}, also commonly used for fitting DSFGs, were investigated. 
However, this model mainly considers the long-wavelength fitting to a modified black body with dust temperatures of $\mathrm{\sim 35 K}$. It does not take into account the near and mid-IR data, 
excluding the PAH contribution. Hence these two models, \citealt{Casey2012} and \citealt{Schreiber2016}, were tested but not used in the results.

The presence of an AGN can break the balance between the absorption and emission from dust, due to their characteristic non-thermal emission. \citet{Fritz2006} 
provides the primary model for the AGN contribution in CIGALE. In terms of the SED fitting, the AGN contribution is mainly an additional mid-infrared dust component. 
\citet{Fritz2006} accurately takes into account the AGN modelling, described by parameters such as the radii and opening angle of the dust torus, and the angle between the AGN axis and line of sight. 
The model is provides the AGN fraction ($\mathrm{ 0 < f_{AGN} < 1}$ where 1 is the 100 \%). 
In this work, we used the AGN model for a broad classification of AGN presence or non-presence in the galaxy. However, the optical depth and the opening angle give the option of a Type 1 AGN (unobscured), 
Type 2 (obscured), an intermediate template or non-AGN (see Table \ref{Cigaletable}). The rest of the values were set to the typical values \citep{Fritz2006} in order to avoid degenerate model templates.
 
\begin{figure}

 \includegraphics[width=1.0\linewidth]{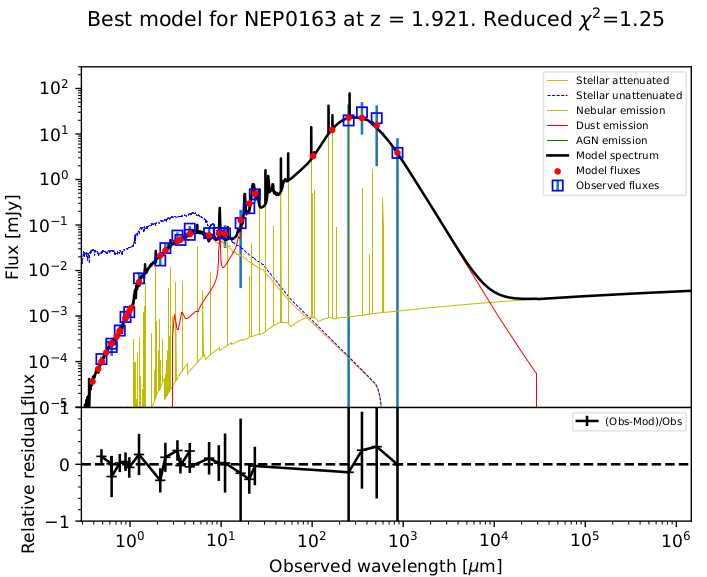} 
 \includegraphics[width=1.0\linewidth]{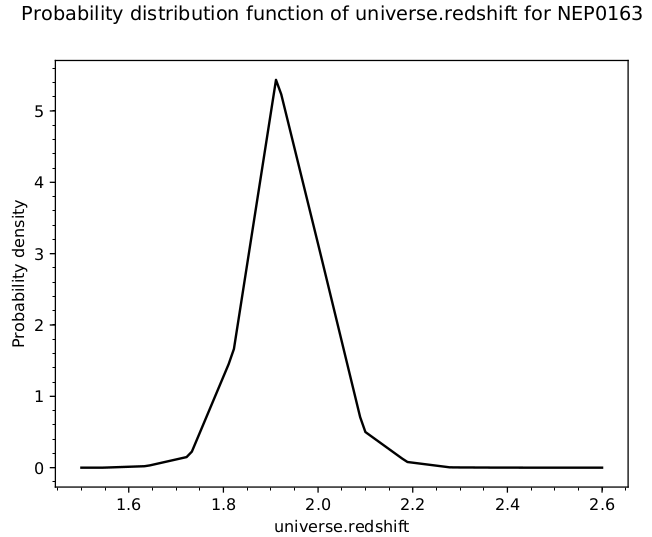}

\caption{ (Left panel) Example galaxy SED fit from our sample produced by CIGALE with the modules taken into account in the legend. The blue squares are the observed fluxes in mJy, 
the red dots represent the corresponding predicted fluxes by CIGALE that lie on the black line model fit.  
\citet{Bruzual2003} reproduces the stellar component,  where the orange line is the stellar attenuation whereas the blue line shows the unattenuated stellar emission. 
The green line is the AGN component, modelled by the \citet{Fritz2006} (no green line implies no AGN presence). 
The dust component is represented by the red line \citep{Draine2014}. The emission lines are shown in pale yellow. 
The upper part of the plot shows the ID of the source, the redshift and the reduced $\mathrm{\chi^{2}}$ fit result. 
(Right panel) Corresponding probability distribution function (PDF) of the estimated photometric redshift for the source in the left panel. 
The PDF peaks at redshift $\mathrm{z_{phot}= 1.9 }$. CIGALE produces a Bayesian analysis for each parameter selecting the most likely result.}
\label{SED_example_PDF}

\end{figure}

The combination of the modules described above, with the grid of selected primary parameters in Table \ref{Cigaletable}, produces $\sim$465 million potential spectral templates for each source. 
CIGALE takes into account those templates and computes the SED fits, selecting the best probabilistic result by a Bayesian approach (see \citealt{Noll2009} for details). 
The best fit SED and probability distribution function (PDF) (see example in Figure \ref{SED_example_PDF} left and right respectively) of each of our high-z candidates were visualised individually 
to remove sources with a possible miss-match, gravitational lenses or an energy balance problem (see Section \ref{Results}). 

An additional quality check is provided by the CIGALE software in the form of a set of estimated flux densities. 
CIGALE produces a mock catalogue of photometry where the flux densities are known from the best fit model. 
This mock photometry can also be used to calculate the exact physical parameters which can then be compared with the estimated values (see \citealt{Boquien2019} for details). 
The reliability of the photometric redshift was validated together with the rest of the physical parameters comparing the real value with the estimated value for each source (see Figure \ref{mock}).

\begin{center} 
\center
  \begin{figure*}
 \includegraphics[width=0.33\linewidth]{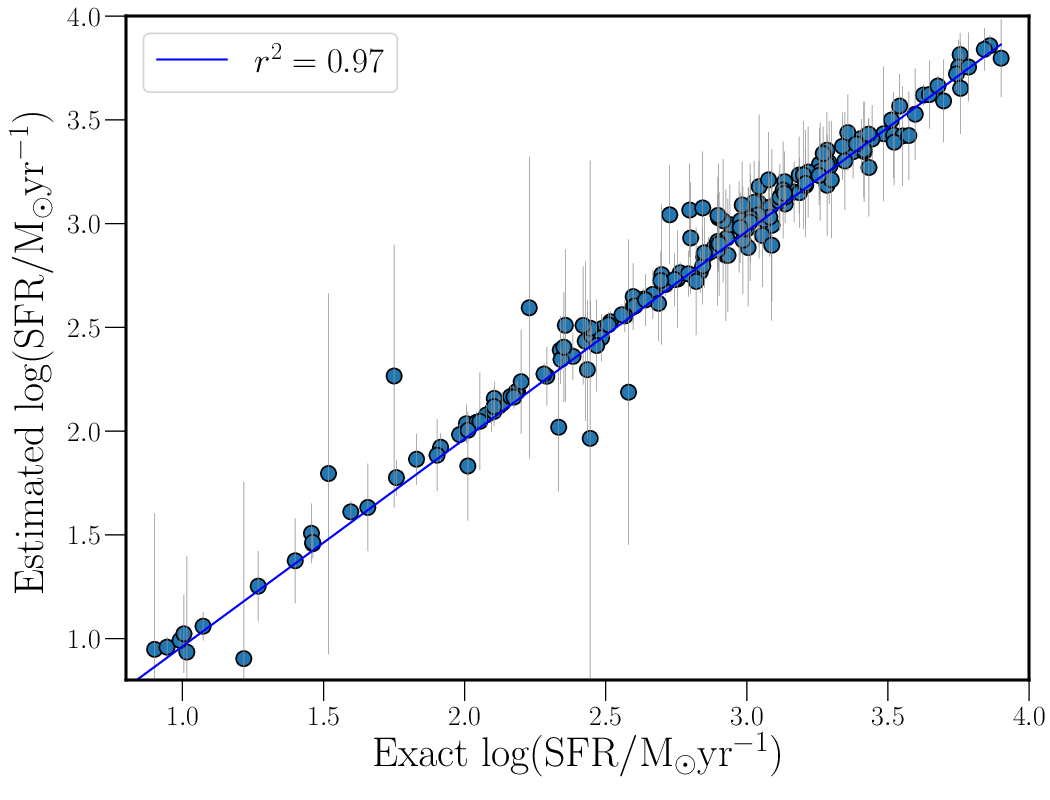}
 \includegraphics[width=0.33\linewidth]{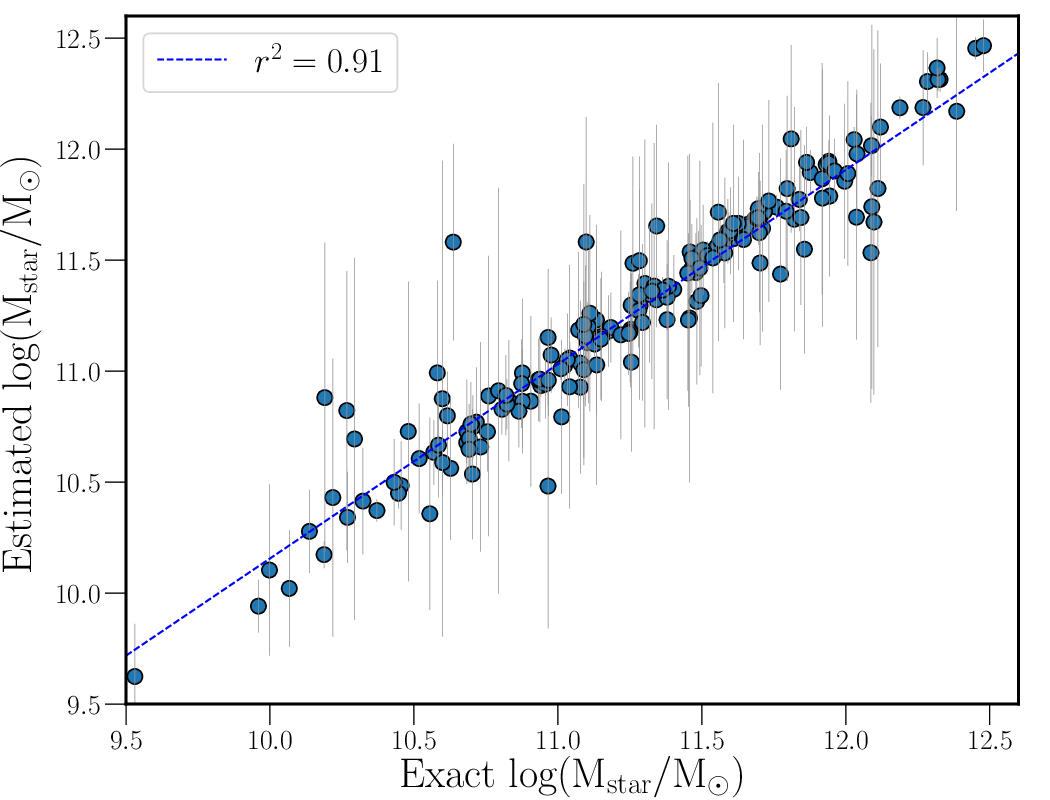} 
 \includegraphics[width=0.33\linewidth]{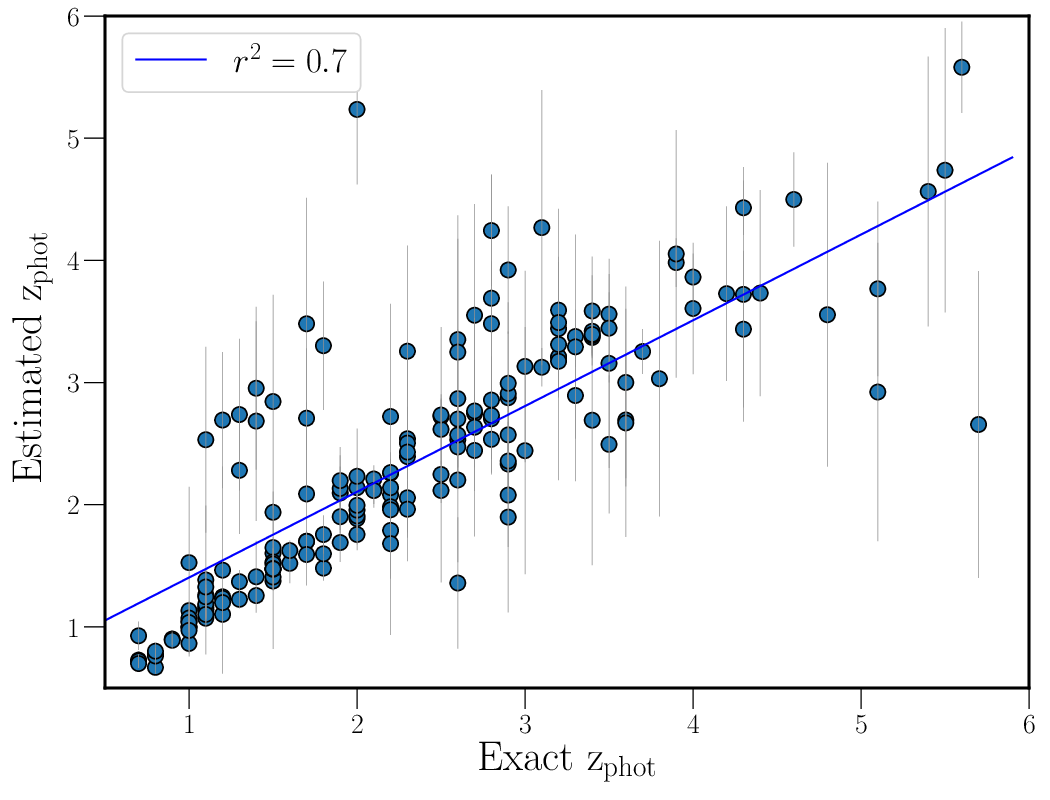} 
 \caption{Quality analysis was produced with the mock photometry to calculate exact values for selected physical parameters: the star formation rate, the stellar mass and the photometric redshift. 
The x-axis represents the exact parameter calculated from the mock photometry from the best fit, the y-axis represents the estimated value presented in the results. 
The blue line is the regression line with the correlation coefficient shown in the top-left of each panel. The good agreement between the two values implies a good reproducibility for the SFR, stellar masses and photometric redshifts.}
 \label{mock}
  \end{figure*}
\end{center}

\section{Results}
\label{Results}

The SED fitting using CIGALE was performed on our high-z sample (Section \ref{Methodology}), 
in order to calculate the physical parameters such as the stellar and gas masses, star formation rates and photometric redshifts. 
Sources with less than seven photometric detections were not included in the following statistical analysis, due to the limitation in calculating reliable stellar masses and SFRs. 
However, this lack of photometric data in the optical-NIR wavebands could also be indicative of their high-z nature, so we do not disregard them out of context. 
The part of these sources with photometric redshifts, $\mathrm{z>3}$  (51 sources, most of them with only SPIRE and IRAC data) are listed in Appendix\label{5detectedsources}. 
In addition, after visually checking the fitted SED, sources that were either probable miss-matches or possible gravitational lens candidates were also removed 
from the statistical sample \citep{Malek2018}. 

We distinguish between the NEP-Wide and NEP-Deep areas during the analysis due to the difference in the number of bands and depth of the AKARI and SPIRE data. In the NEP-Deep field, 
88\% of the galaxies have more than twelve photometric detections, and on average detections in more than $\mathrm{\sim}$ 20 bands. In the NEP-Wide field, 
the number of sources with more than ten photometric detections is 60\% with on average detection in 11 photometric bands. 

The final sub-sample with good multi-wavelength coverage and discarding mis-matches contains 185 high-z dusty galaxies: 78 in the NEP-Deep field and 107 in the NEP-Wide field.

%
\subsection{Photometric redshift}
\label{photometricredshiftresults}

Photometric redshifts are often used as a proxy in large surveys due to the lack of available spectroscopic redshifts which are observational time-consuming to collect. 
In particular for DSFGs, the high amounts of dust present at high redshift makes redshifts from optical lines impossible to obtain, and the use of photometric redshift is the only viable route for 
large sample of galaxies \citep[see e.g.][]{daCunha2015}.  

We used CIGALE to calculate the photometric redshift for the final selection of 185 high-z galaxies with good photometric coverage. 
The photometric redshift - and consequently, the rest of the physical parameters, depend on the models assumed in CIGALE. 
CIGALE takes into account the entire spectral range to calculate the photometric redshift. Hence, the selection of the dust emission model can produce differences in the photometric redshifts. 

The average redshift for the entire high-z DSFG sample, independent of the selection method (SPIRE colours and SCUBA-2 flux, was found to be 
$\mathrm{z= 2.33 \pm 0.08}$ ($\mathrm{0.1 < z_{phot} < 5.6}$).  
The redshifts for the sources selected with SCUBA-2 fluxes is on average lower than the sources selected via SPIRE colours. 
The latter sample peaks at a higher average redshift with median $\mathrm{2.57^{+0.08}_{-0.09}}$,  
whereas the SMGs have a median redshift  of $\mathrm{1.48^{+0.2}_{-0.06}}$.  
The redshift distribution of the two populations is shown in Figure \ref{histzphot}. The redshift distribution over the NEP-Wide and NEP-Deep fields were found 
to be similar and are represented together in the histogram in Figure \ref{histzphot}. 
The higher redshift nature of the SPIRE selected DSFGs indicates that the 500$\mathrm{\, \mu m}$ riser method and SPIRE colour-colour diagrams 
select on average higher redshift sources than with SCUBA-2 fluxes, as expected. 
There was not found to be any dependence of the photometric redshift  on the number of photometric bands.  

We also investigated the spectroscopic data sets available over the NEP field (\citet{Shim2013}, Kim et al. in prep., see Section \ref{specredshift}). 
Unfortunately, spectroscopic data is only available for 10 of our sources in our high-z catalogue (4 for the SCUBA-2 selected galaxies and 6 for the SPIRE selected galaxies, that is $<$ 5\% of the sample). 
This is not a large enough sample to extract any conclusions on the quality of the photometric redshift estimate. For the handful of spectroscopic redshifts that are available, 
these were obtained from optical and NIR lines and all have low-quality flags in the spectroscopic catalogues. Therefore, to confirm the quality of the photometric redshift, 
spectroscopic data from sub-millimetre telescopes is proposed for future work (see Section \ref{conclusions}).

\begin{center} 
\center
  \begin{figure} 
   \includegraphics[width=\columnwidth]{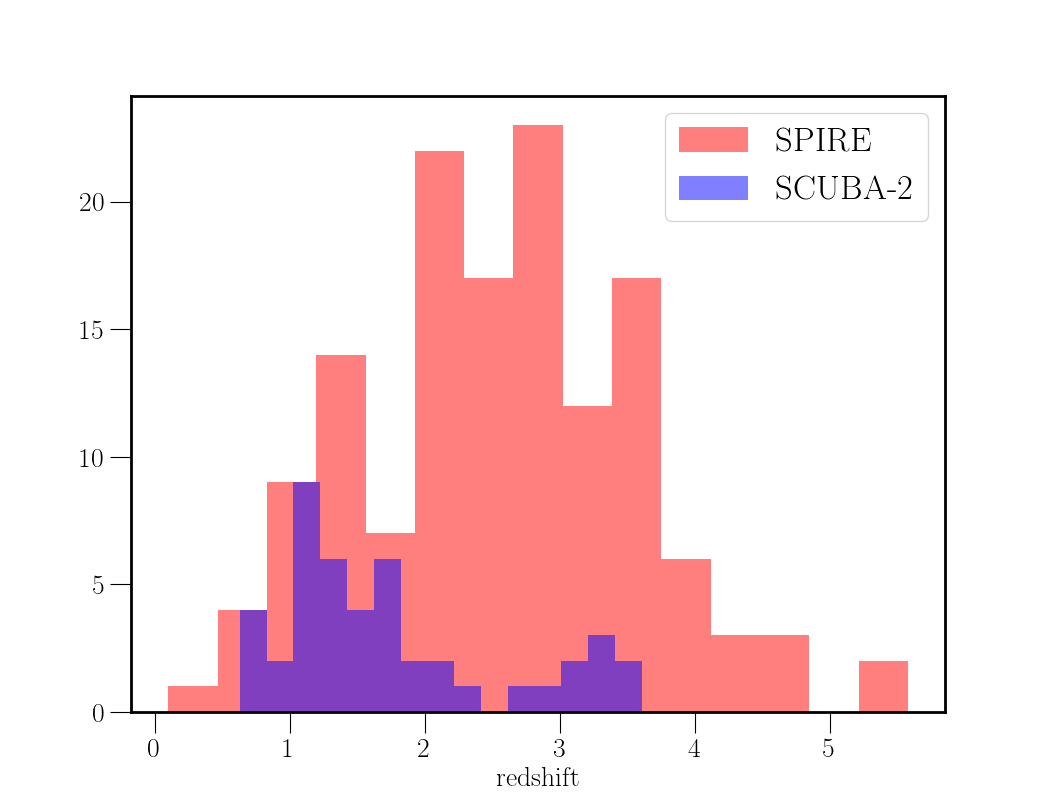}
 \caption{ Histogram of the calculated photometric redshifts for the two selection methods: sample selected with SCUBA-2 fluxes (blue) and  selected with SPIRE colours (red). 
 The sample selected via SPIRE colours peaks at a higher redshift with a median value of $\mathrm{2.57^{+0.08}_{-0.09}}$, whereas the SCUBA-2 selected 
 sources have a lower median redshift of $\mathrm{1.48^{+0.2}_{-0.06}}$. }  
 \label{histzphot}
  \end{figure}
  \end{center}

\begin{center} 
\center
  \begin{figure} 
    \includegraphics[width=\columnwidth]{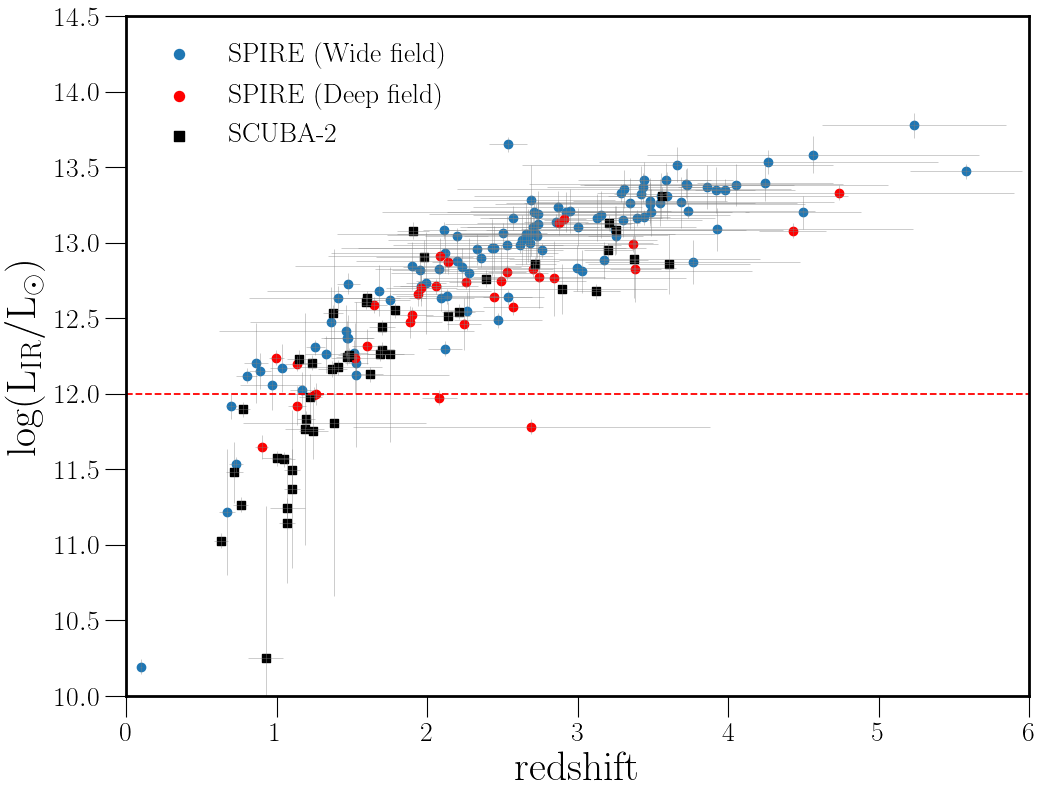}
   \caption{Infrared luminosity ($\mathrm{L_{IR}}$) against photometric redshift for the NEP-Deep field (black and red dots) and the Wide field (blue dots), 
   with the selection method detailed in the legend. 
   The red dashed line shows the limit for ULIRGs $\mathrm{L_{IR} > 10^{12} L_{\odot}}$: above which the sources are either ULIRGs (50\%) or hyperluminous infrared galaxies 
   (HLIRGs; $\mathrm{L_{IR} > 10^{13} L_{\odot}}$) (36\%). 
   The remaining sources below the line (14\%) are  classified as luminous infrared galaxies (LIRGs, $\mathrm{ 10^{11} L_{\odot} < L_{IR} < 10^{12} L_{\odot}}$ )
   The galaxies selected with the SCUBA-2 method have the same number of photometric bands as the sources selected with SPIRE in the NEP-Deep field, however, the 
   SPIRE colour criteria select, on average, more luminous sources (15 \% more) than galaxies selected with SCUBA-2.} 
 \label{LIR_vs_z}
  \end{figure}
  \end{center}
 Using the photometric redshift, CIGALE calculates the infrared luminosity. Figure~ \ref{LIR_vs_z} shows the resulting luminosity-redshift distribution for each of our selection criteria. 
 The SPIRE selected sources are further split into sources in the NEP-Deep and Wide fields. 
 The calculated luminosity in the NEP-Wide and NEP-Deep field is similar over the same redshift range, 
 and is therefore not biased by the depth of the data. 
 From Figure~ \ref{LIR_vs_z}, it can be seen that almost all of the sources are ULIRGs, as expected, and  with $\mathrm{z < 1}$ are mostly sources selected with SCUBA-2 flux method. 
 The detection limit reached for SPIRE at the shortest wavelength is $\mathrm{\sim 10 mJy}$ whereas for the SCUBA-2 is $\sim 3 mJy$.  
 
Due to the larger areal coverage, there are four times more luminous galaxies at $\mathrm{z> 3.5}$ in the NEP-Wide than in the NEP-Deep field. 
However, the galaxies selected with the SCUBA-2 method are 15 \% less luminous on average than the galaxies selected via the SPIRE colours over the same area with the same number of bands. 
It appears that SPIRE colour and 500$\mathrm{\, \mu m}$ techniques tend to select more luminous galaxies, probably because they are more massive and warmer (see Section \ref{SFR_MS}).
 
A total of 12 sources are detected at z $\gtrsim$ 4 from both selection methods combined. 
This number can be compared with the surface density predicted by the contemporary galaxy evolution models of \citet{Pearson2017}. 
Over a similar area, \citet{Pearson2017} predicts around 20 sources above the SPIRE confusion limit at z $>$ 4. 
Taking into account the fact that some of our high-z sources were rejected due 
to the low number of photometric detections (10 of the rejected list have z$\gtrsim$4, see Table~\ref{List_rejecteddetections} in  Appendix~\ref{appendix_highz}), 
the predictions are broadly consistent with the numbers presented here. 

%
\subsection{The star formation rate and the main sequence of galaxies}
\label{SFR_MS}

The star formation rate (SFR) and the stellar mass were computed to determine the position of our high-z sources on the main sequence (MS) of galaxies, 
to address whether DSFGs selected at sub-millimetre wavelengths lie on the MS or above it as outlying starbursts.

The median star formation rate for the entire sample of 185 DSFGs is $\mathrm{SFR= 797^{+108}_{50} \ M_{\odot} yr^{-1} }$.  
Over the NEP-Deep field we see a clear distinction between sources selected using  SCUBA-2 flux ($\mathrm{SFR= 146^{+32}_{-19} \ M_{\odot} yr^{-1} }$) 
and SPIRE colour ($\mathrm{SFR= 444^{+84}_{-55} \ M_{\odot} yr^{-1} }$) 
criteria. The SCUBA-2 flux method is able to select relatively faint sources with a more moderate star formation rate than the 
extreme sources detected with SPIRE colours  \citep[see e.g.][]{Riechers2013, Rowan-Robinson2016, Rowan-Robinson2018}. The extremest source we find have $\mathrm{SFRs < 7,000 M_{\odot yr^{-1}}}$
The median SFR over the NEP-Wide field is higher ($\mathrm{SFR= 1265^{+138}_{-92}  \ M_{\odot} yr^{-1} }$), however, both fields exhibit more moderate SFRs than reported in previous work 
\citep[e.g.][see Section \ref{Discussion} for discussion]{Rowan-Robinson2016}. 

We calculated the stellar mass for our sample using CIGALE. Two different stellar population models were tested and the SED fitting was run for both the \citet{Bruzual2003} 
and \citet{Maraston2005} models supported by CIGALE. 
No significant differences in the results were found that would change the position of the sample relative to  the MS of galaxies.

We conclude that the enormous differences in submillimetre galaxies discussed in the literature \citep[see e.g.][]{daCunha2015, Dunlop2017} 
are not influenced by using stellar populations models based on isochrones such as \citet{Bruzual2003} instead of turn-off based models such as \citet{Maraston2005}. 
For this work, the models of  \citet{Bruzual2003} were finally chosen in order for us to compare our results with the work in the literature that uses the  alternative MAGPHYS SED fitting code, 
which uses the same population model \citep[e.g.][]{daCunha2015, Miettinen2017a}. 

The main sequence of galaxies is known to evolve with redshift and for this work, the definition of \citet{Speagle2014} was applied to
\begin{equation}
\mathrm{log(SFR)= 0.84 }$ $\mathrm{ - 0.026t_{z} log(M_{*} + 0.11t_{z} -6.51)}
  \end{equation}
where $\mathrm{t_{z}}$ is the age of the Universe in [Gyr]. Note that the definition is function of redshift and we considered the MS at the corresponding redshift for each galaxy. 
A galaxy is commonly considered 'above the MS' if its position is a multiple of times above the definition of the MS line. 
The correlation is not strictly a line and has some scatter which makes considere galaxies above the MS if their SFR is several times above the line. 
\citet{daCunha2015} uses a criteria of three times above the MS, although others assume different criteria, such as \citet{Elbaz2011}, 
who defines the MS as a function of cosmic time and considers the sources two times above the MS as outlying starburst galaxies.  
Here, in order to compare our results with \citet{daCunha2015} and \citet{Dunlop2017}, 
who assume the same definition of the MS as \citet{Speagle2014}, 
all sources that are three times above the line of the MS are considered to be above the MS and therefore to be starburst galaxies (see Figure \ref{MS_bins}). 

We find that 38\% of our sources would be defined as starbursts lying three times above the MS.  
The remaining sources being on the MS despite of their high SFRs because they are massive galaxies, which 
can be visualised by computing the specific star formation rate $\mathrm{sSFR = SFR / M_{star}}$ (see top-panel of Figure \ref{sSFR_vs_redshift}). 
The bottom-panel of Figure \ref{sSFR_vs_redshift} 
shows the number-redshift distribution of the starburst and main sequence galaxies. Most starburst galaxies are found at z=2-4 with activity descreasing to higher redshift 
$\mathrm{z > 4}$ (however, this decrease is accompanied by a decrease in the sample that makes difficult to be certain of evolution of the starburst activity with redshift). 

\begin{center} 
\center
  \begin{figure} 
    \includegraphics[width=\columnwidth]{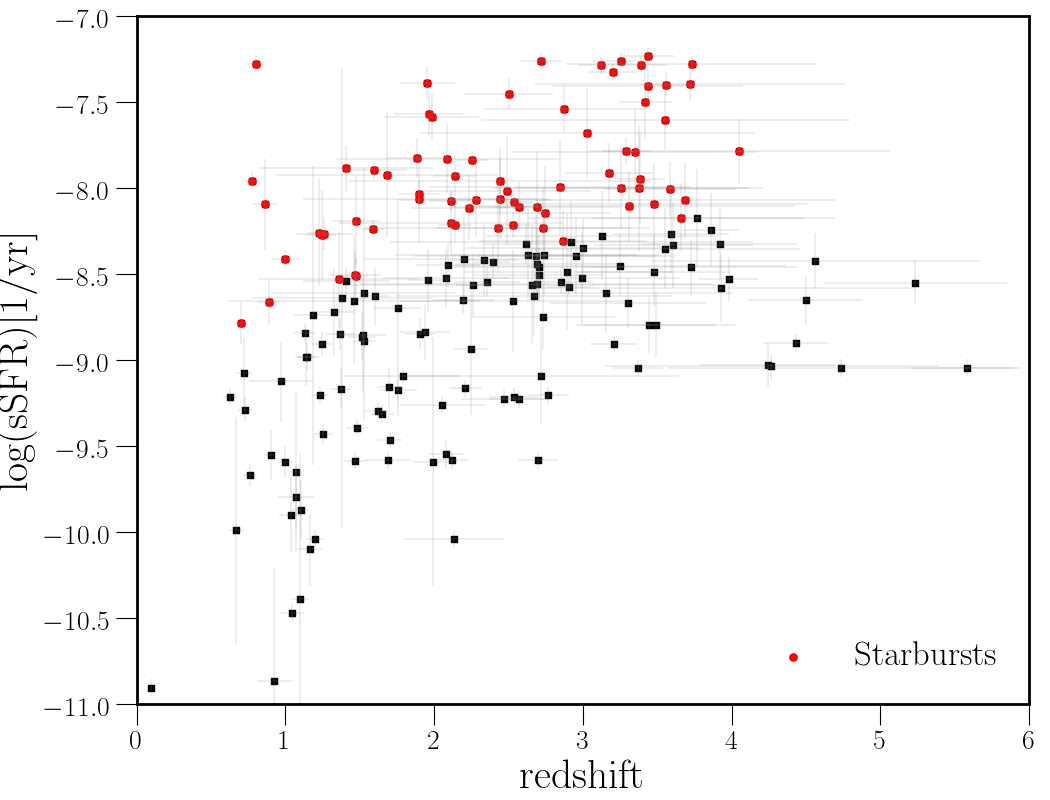}
    \includegraphics[width=\columnwidth]{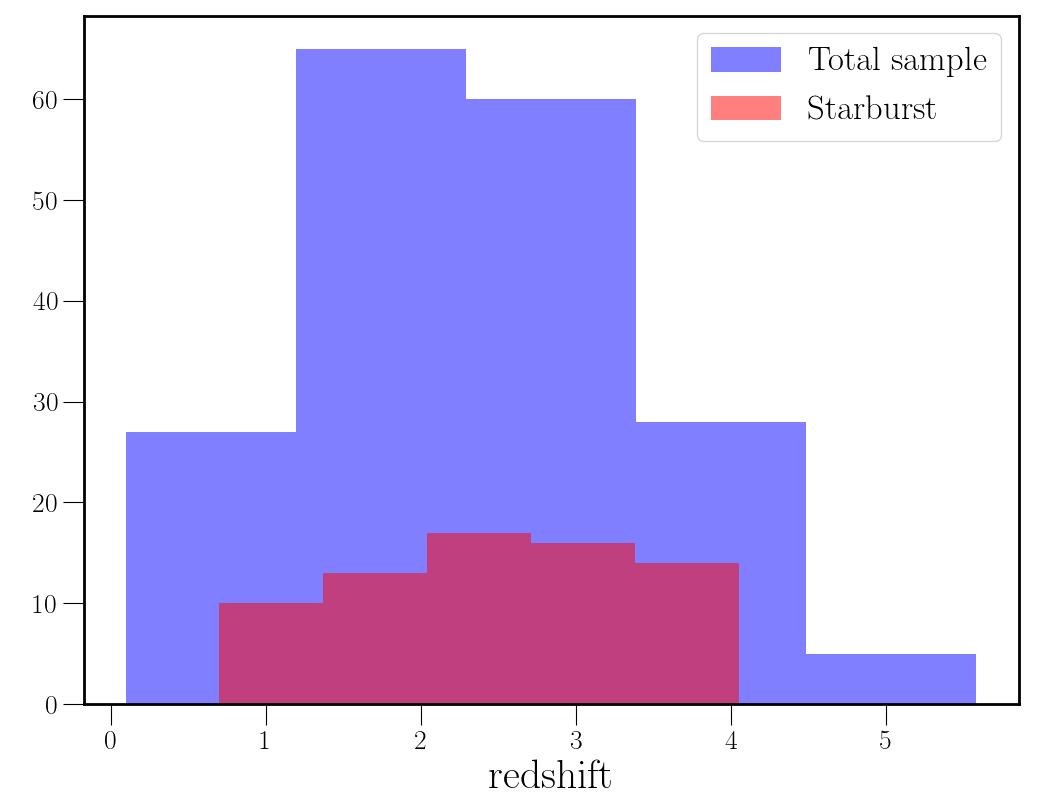}
   \caption{(Top panel) Specific star-formation rate (sSFR) against redshift for the total sample. 
   The galaxies classified as starbursts using the definition of  \citet{Speagle2014} are marked as red dots and are seen to exhibit a higher sSFR than the rest of the population. 
   (Bottom panel) Number-redshift distribution of the total sample (blue bars) and the starbursts (red bars). 
   The fraction of starbursts increases up to z=2, and it is similar out to  $\mathrm{z \sim 3-4}$ (around 40\%), decreasing to higher redshifts $\mathrm{z > 4}$. 
   A larger sample at $\mathrm{z > 4}$ would be required to investigate the evolution of the starbursts over cosmic time.} 
 \label{sSFR_vs_redshift}
  \end{figure}
  \end{center} 

In order to evaluate their position in the MS and any evolution with cosmic time, the sample 
was segregated into four redshift bins ($\mathrm{1.5 \leq z < 2.5}$, $\mathrm{2.5 \leq z < 3.5}$, $\mathrm{3.5 \leq z < 4.5}$  and $\mathrm{4.5 \leq z < 5.7}$).
Figure~\ref{MS_bins} shows the main sequence for the redshift bins centred on the redshifts  $\mathrm{z = 2}$ and $\mathrm{z = 3}$ respectively.
The fraction of starburst galaxies is highest in the $\mathrm{z=2}$ redshift bin (43\%), decreasing slightly to $\mathrm{z=3}$ (40\%) over a similar size sample. 
The rest of the redshift bins are too sparsely populated to draw robust conclusions on any evolution of the MS with redshift in our sample and are not shown in the figure. 
We further compared the position on the MS for our two selection methods,  over the NEP-Deep field, where there are the same number of bands, and similar depth.  
We find 12 \% more galaxies classified as starbursts using the SPIRE colours compared to the SCUBA-2 selected sources.

\begin{center}
\center
  \begin{figure*}
   \includegraphics[width=\columnwidth]{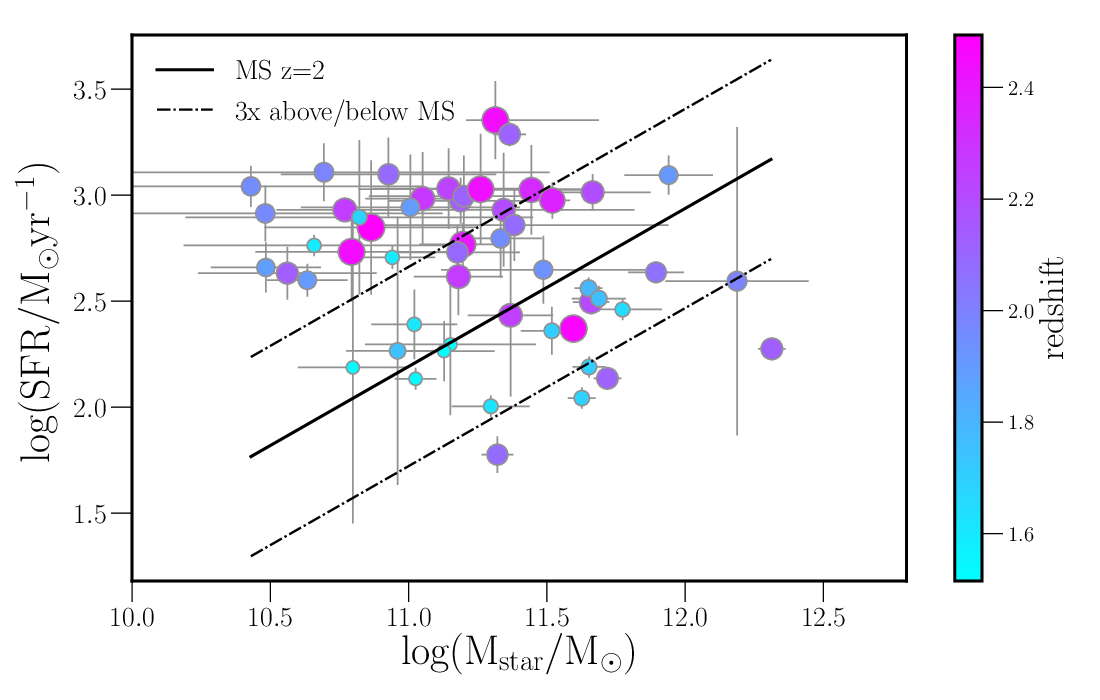} 
      \includegraphics[width=\columnwidth]{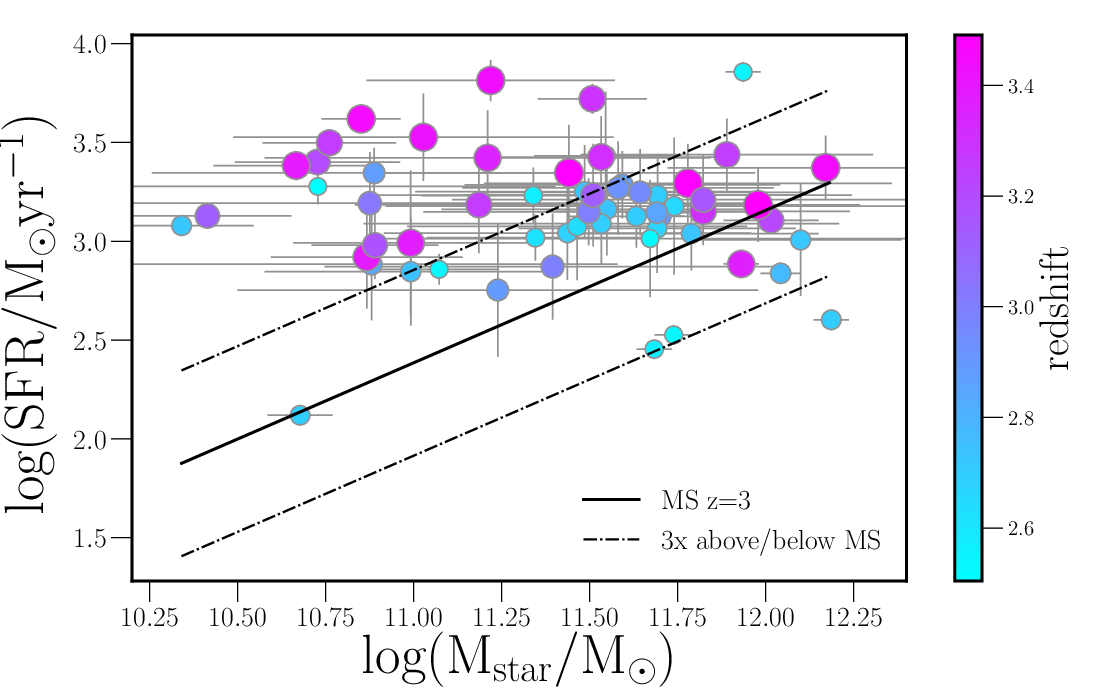} 
\caption{Main sequence of galaxies. The black line shows the MS as defined in \citet{Speagle2014}, 
with the dashed lines showing three times above and below the line of the MS for this redshift bin. The sources considered as starburst are three times above the MS. 
Left panel: The main sequence of galaxies for the redshift bin at 
$\mathrm{1.5 \leq z < 2.5}$. In the $\mathrm{z=2}$ bin the fraction of starbursts with respect to the number of sources is 43\%. 
Right panel: The main sequence of galaxies for the redshift bin at $\mathrm{2.5 \leq z < 3.5}$. In the $\mathrm{z=3}$ bin the fraction 
of starbursts with respect to the number of sources is 40\%. In both cases the starbursts tend to be of lower stellar mass than the majority of sources lying on the MS.} 
 \label{MS_bins}
 \end{figure*}
 \end{center}

\subsection{The effect of AGN}
\label{SFR_AGNPAH}

\begin{figure} [h!] 
  \includegraphics[width=\columnwidth]{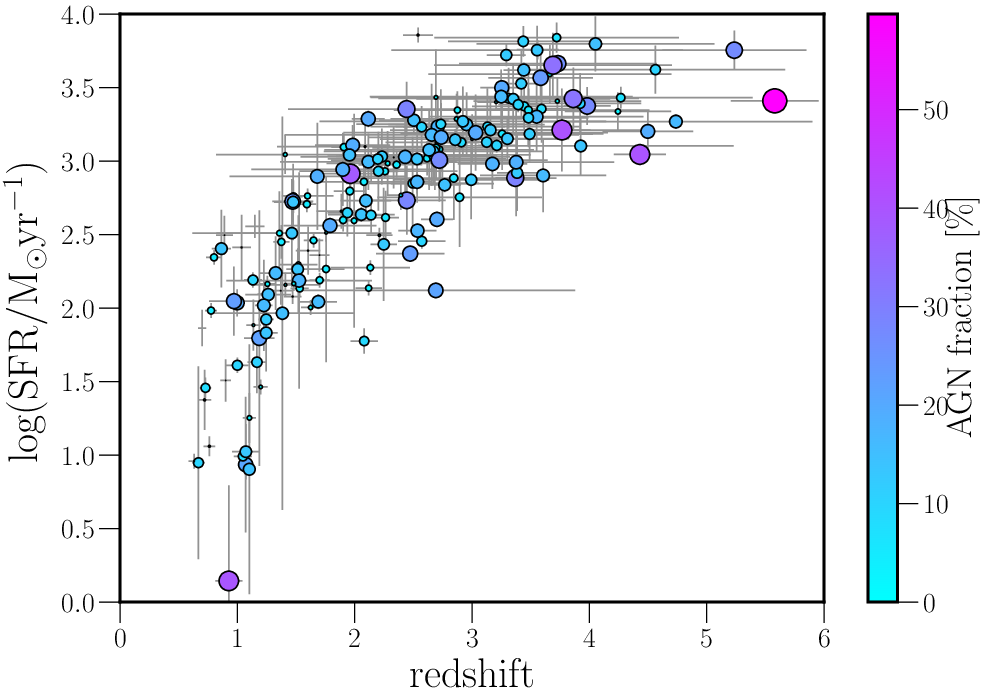} 
\caption{ SFR against redshift with the AGN fraction in the colour bar: 12\% of the high-z sample have an AGN presence of more than 20\% with no correlation with redshift. } 
\label{SFR_AGN} 
\end{figure}

We also investigated the effect of any AGN presence and associated PAH deficit \citep{Elbaz2011}, 
both significant parameters that could affect the star formation rate, thus influencing the positions of sources on or off the MS. 
The photometric bands allowing to probe the PAHs contribution and AGN presence in mid-IR are the AKARI bands (S9, S11, L15, L18, L24), and the Spitzer (MIPS1) 
(see Section \ref{observations}). 
We use CIGALE to estimate any possible AGN contribution to a spectral fit assuming the AGN model of \citep{Fritz2006}. 
Approximately 12\% of the high-z sample have an AGN presence of more than 20\%, with a handful of sources classified as AGN dominated. 
We do not find a correlation between AGN presence and SFR or redshift (see Figure \ref{SFR_AGN}), neither 
with a high AGN fraction being specifically associated with the starburst galaxies above the MS. 
However, any AGN presence will affect the estimated SFR (of the order of $<$10\% for the majority of our sample) and it is necessary to take an AGN model to the study of 
DSFGs. 

The contribution from PAHs concentration was evaluated using the \citealt{Draine2014} dust models (see Table \ref{Cigaletable}). 
We do not find any PAH deficit in the starburst galaxies having a similar fraction of PAHs for galaxies above or on the MS. 
The sources that show a high AGN presence have small or non-PAH fractions as expected due to grain destruction by the intense radiation from the AGN , 
especially at the highest redshifts ($\mathrm{z > 4}$).

%
\subsection{Star formation efficiency}
\label{SFR_SFE}

The star formation efficiency (SFE) relates the star formation rate to the gas mass ($\mathrm{SFE = SFR/M_{gas}}$) and is a measure of how efficiently the gas is processed into stars. 
The SFE allows us to analyse the factors that trigger the star formation. 
We compared the SFE of the populations both above and on the MS in our sample to see if there is a factor 
that can trigger the star formation or if it is simply the fact that they contain more gas. 

The mass of the gas is calculated by applying a conversion factor for the dust-to-gas ratio, to the mass of the dust derived from the \citet{Draine2014} models. 
There is evidence than high-z submillimetre galaxies are dust rich, however, the high dust content is proportional to the high gas content, having a dust-to-gas ratio higher or similar to normal spiral galaxies \citep{Santini2010}.

\begin{figure}
     \includegraphics[width=1.0\linewidth]{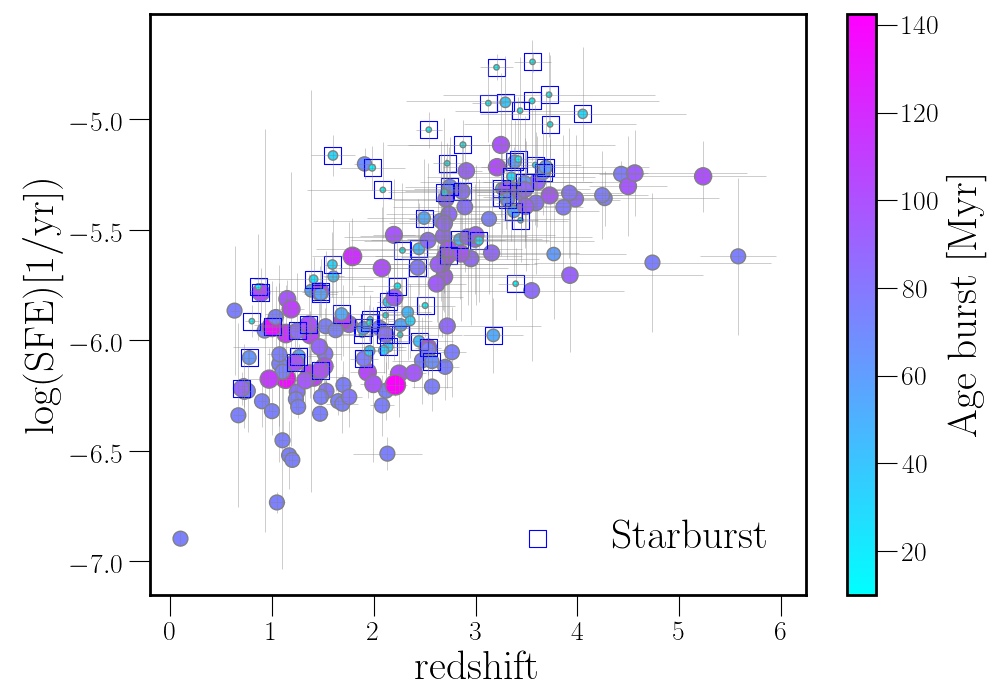}
      \includegraphics[width=1.0\linewidth]{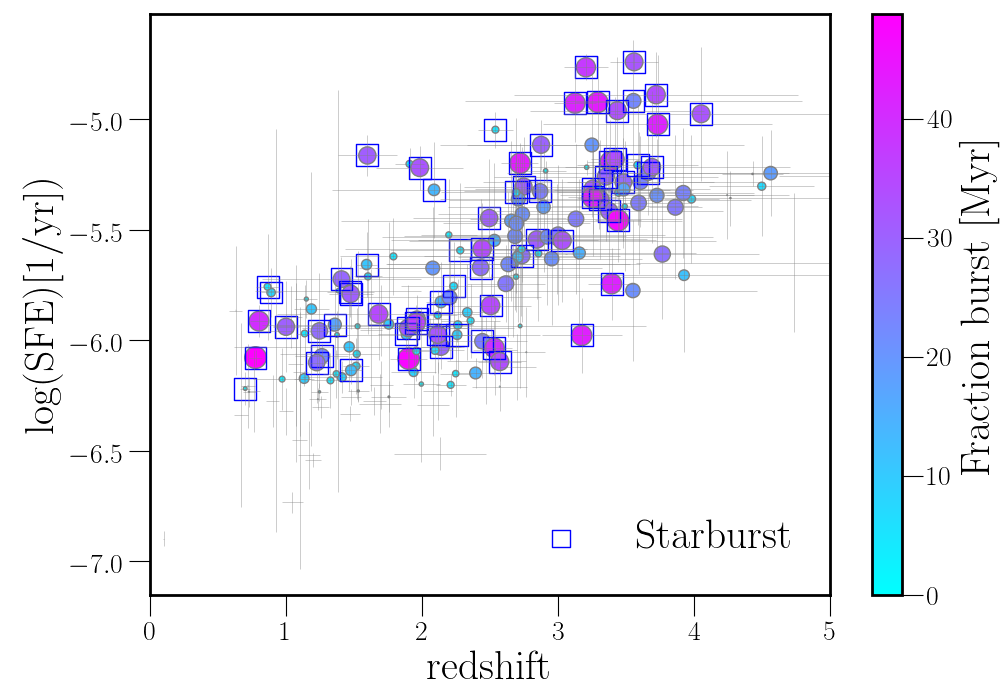}
   \caption{ SFE as a function of redshift. A general trend of increasing SFE with redshift is seen. 
   SFE as a function of redshift. A general trend of increasing SFE with redshift is seen. 
   This trend is more prominent in the starburst population (above the MS) identified by the blue border squares around the circles in the plot. 
   Top-panel shows the dependence of the results on the age of the starburst (in the colour-bar). 
   There is a clear dependence between the efficiency of star formation with the age of the starbursts: 
   the later the starburst in the age of the galaxy the more efficient the star formation. 
   The bottom-panel shows the same SFE-redshift plot with the dependence on the fraction of the star-formation burst (in the colour-bar), the more significant burst, 
   the more efficient the star formation. }
 \label{SFE_frac}
  \end{figure}

In Figure \ref{SFE_frac}, the SFE is plotted as a function of redshift for our entire high-z sample. The SFE correlates with redshift, being the sources at higher redshift being more efficient at forming stars. Also shown in Figure \ref{SFE_frac} (top-panel colour bar) is the age of the burst of star formation in the galaxy 
as derived from CIGALE (as described in Section~\ref{Methodology} and Figure~\ref{SFH_2exp_starburst}). 
The age of the burst in the galaxy corresponds to the moment that the burst occurs in the time of the galaxy. 
We find a clear trend between the age of the starburst in the galaxy and the efficiency of the star formation: the later the burst is in the age of the galaxy, the more efficient the star formation. 
  
The age can be a difficult parameter to calculate thought SED fitting modelling \citep{Buat2014}. Also, the mock test shows a bad estimation for the age of the burst with a poor linear regression ($\mathrm{r^{2}=0.3}$). For this reason, we also evaluate the presence of any starburst by the parameter $\mathrm{f_{burst}}$ 
with a better result in the mock catalogue ($\mathrm{r^{2}=0.7}$) than the age of the burst. The starburst fraction has an effect on the SFE of the galaxy, the more massive the starburst, the bigger the triggering of star 
formation (see Figure \ref{SFE_frac} bottom-panel). 
  
We find no relation between either the AGN fraction or the PAH concentration with the SFE, 
which was expected from the previous lack of correlation between those parameters with the star formation discussed in Section~\ref{SFR_AGNPAH}.

The presence of a starburst could be due to a merger. To confirm this, the morphology of the source would be required. As future work, 
the SFE will be calculated by tracing the molecular gas via spectroscopic observations to study the compactness of the sample. 
Furthermore, the SFH parameters are challenging to constrain from broadband SED fitting and spectroscopic observations will help to better calculate these parameters (see future work in \ref{conclusions}). 
 
%
%
\section{Discussion}
\label{Discussion} 

There is no explicit consensus on the nature of sub-millimetre galaxies over cosmic time. In this Section, we discuss the reasons for the differences 
in our results with previous studies and give our vision of the nature of sub-millimetre galaxies, discussing their position on the MS, shedding some light on the considerable disagreement 
in the literature, explaining if the difference relies on selection effects or model considerations.   

%
\subsection{High-z selection at submillimetre wavelengths} 

We studied the high-z population at the NEP by using two different sub-millimetre selection criteria: SCUBA-2 selected sources at $\mathrm{850 \mu m}$ and selection by SPIRE 
colours including the 500$\mathrm{\, \mu m}$ riser technique. This approach allows us to evaluate if the different results in previous works could be due to a selection effect. 
The SCUBA-2 selected sample has a significantly lower fraction lying above the MS compared to the SPIRE colour selected sources at shorter wavelengths (see Section \ref{Results}). 
The SCUBA-2 methods appear to select sources at lower IR luminosity than the SPIRE selection method over the same NEP-Deep field and, in consequence, lower star formation. 
Such a difference in SFR due to the different population could influence the different outcomes in literature, but it is not big enough to be the main reason for the disagreement. 
We find lower star formation rates than \citealt{Rowan-Robinson2016} using the same selection method, 
which implies that these differences are not a selection effect. These extreme star-forming galaxies are 'monsters' 
with large stellar masses and SFRs. However, both the consideration of an AGN presence and avoiding fitting with local galaxy templates (e.g. Arp220, M82, \citealt{Polletta2007}), 
it is possible to lower the estimated SFRs by an order of magnitude. 

%
\subsection{The position of high-z SMGs on the MS}
We find that the position of galaxies on the MS relation cannot only be explained by selection effects. 
Assumptions in the modelling approach, such as the different physical models in the SED fitting, can affect the calculation of the SFR and stellar masses. 
We investigated the differences from assuming stellar population models by using the models of \citet{Bruzual2003} and \citet{Maraston2005}. 
Although there were small differences in the estimated SFR and stellar masses, these differences were not enough to change the position of the sub-millimetre galaxies in the MS diagram significantly.  

Our results are in disagreement with some previous studies such as \citet{Ikarashi2017}, where 72\% of their submillimetre selected sample lie above the MS 
with $\mathrm{1.4 \leq z < 2.5}$, or with \citet{Miettinen2017a} who measures 63\% of SMGs above the MS. 
Some of this discrepancy could be since we take into account the AGN contribution in the fitting, which imply lower SFRs, and the differences in the code assumptions (we use CIGALE instead of MAGPHYS). The AGN emission has an influence on the MS slope up to 32 \% depending on the SFH and the type of AGN \citep{Ciesla2015}. Note that in particular, the MAGPHYS SED fitting code uses the \citet{Bruzual2003} models, which means that we should find similar stellar masses with CIGALE. 
However, SFRs may differ due to the different SFH and the lack of AGN models. Hence, with similar stellar masses and different SFRs, it will change the position in the MS. 

Other submillimetre samples, such as \citet{Dunlop2017}, suggest that all their sources lie on the main sequence. This disagreement with 
our results may indeed be a selection effect since \citet{Dunlop2017} is a deep pencil-beam survey, and it is sensitive to lower SFRs but comparable stellar masses. 

Another reason related to the modelling itself is the use of local templates for galaxies to calculate photometric redshifts, which tend to give higher redshifts 
\citep{Ma2019} and higher SFRs. Finally, assumed the SFH could produce a different SFR. 

Consequently, it is not surprising that our results are an average between two of the most extreme cases in literature at $\mathrm{z \sim 2}$: \citet{Dunlop2017} 
with no galaxies above the MS and studies like \citet{Ikarashi2017}, \citet{Miettinen2017a}. \citealt{daCunha2015} finds 52\% of the SMGs above the MS at the bin z=2, whereas we find 43 \%. 
Hence, the difference could be due to the use of an AGN model since the AGN presence affects the estimated SFR up to 10\% for the majority of our sample. 

Ideally, the same sample of DSFGs, with spectroscopic redshifts, should be investigated using a set of available SED fitting codes, i.e. MAGPHYS, 
CIGALE and LePhare, quantifying the differences of SFRs between these three codes.

\subsection{SFE of starburst}
The SFE relates the star formation to the gas reservoir within the galaxy. Thus, finding some parameter that correlates with the SFE can also provide a link to the triggering, 
or quenching, of the star formation itself. We find a relation between the presence of a starburst in the galaxy with the SFE. A merger could cause the presence of such starburst episodes, 
and the age of this starbursts could, in turn, be related to the state of the merger. 

The morphology of the galaxies can be related to mergers, which in turn influences the SFR \citep{Elbaz2018}, this also includes the compactness of a galaxy \citep{Elbaz2007}. 
Moreover, the time that the merger occurs in the life of the galaxy can influence in the SFE. 

The bimodality of star formation in SMGs was evaluated by  \citealt{Elbaz2018}: a more compact group with a moderate mode of star formation (on the MS) and another with extreme 
star formation mode (above the MS) with enhanced gas fractions. Spectroscopic observations are required to investigate any relation with the morphology or the compactness of the galaxy. 
For instance, whether the starburst galaxies are more compact than galaxies that lie in the MS \citep{Elbaz2011} and how mergers may enhance the star formation at early stages \citep{Riechers2017}. 
Spectroscopic observations of molecular gas are the best to approach to this matter.


\section{Summary and conclusions} 
\label{conclusions}

Our multi-wavelength approach in deep fields has allowed us to analyse a large sample of DSFGs across their entire spectrum. 
We have utilised two methods to select a sample of high-z candidates. Our analysis finds that:

\begin{itemize}
  
 \item Sources selected using a criterion of SPIRE colours and 500$\mathrm{\, \mu m}$ risers, select more extreme sources than the method using SCUBA-2 fluxes.  The median redshift of the SPIRE selected sample  is $\mathrm{z = 2.57^{+0.08}_{-0.09}}$ whereas the SCUBA-2 selected sample has a median of $\mathrm{z = 1.48^{+0.2}_{-0.06}}$. The luminosities and SFRs differ but, even in the more extreme cases, 
 the SFRs are still only of the order of hundreds or a few thousands of solar masses per year.
   
 \item Although we do not find a correlation between the star formation and AGN presence,  AGN models must be taken into account for the study of sub-millimetre galaxies. 
The luminosities and SFRs are lower when AGN models are considered, which is not frequent in the study of these populations, probably due to the limitation of the SED fitting codes. 
 
 \item Although the majority of our sub-millimetre galaxies generally lie on the MS we find a significant fraction (43\% at $\mathrm{z=2}$) 
 that are classified as starbursts above the MS. 
 
 \item The SFE depends on the epoch and intensity of the starburst in the galaxy, the later the burst, the more intense the star formation. However, they do not strongly depend on the presence of any AGN, 
 which suggests that the trigger of star formation could be related to mergers instead of secular processes.  As future work, spectroscopic observations with sub-millimetre telescopes will be carried out to confirm it. 
\end{itemize}

We conclude that the sub-millimetre population is a mix of different galaxies and, that by selecting sources based on either their SPIRE colours or a SCUBA-2 detection, their properties vary. 
We conclude although most sub-millimetre galaxies (60\%) lie on the MS, that there is a population of starbursting galaxies of extreme nature. However,  galaxies such as HFLS3 \citep{Riechers2013} are likely not to be representative of the general sub-millimetre population; 
nonetheless, more galaxies with extreme star formation could be found by the proposed selection methods.  We conclude that a multi-wavelength approach in deep fields - with comprehensive, wide multi-wavelength photometric coverage (such as the NEP region) - is necessary for a better understanding of this population. 

This study requires spectroscopic confirmation for further evaluate the SFE which will be subject of future work by carrying out observations using facilities such as the Large Millimeter Telescope (LMT) \citep{Hughes2010}. 
    
\begin{acknowledgements}
We thank the anonymous referee for a thorough and constructive report.
LB is very grateful for the support of the ESA Research Fellowship.
KM has been also supported by the National Science Centre (grant UMO-2018/30/E/ST9/00082). 
GJW gratefully thanks The Leverhulme Trust for funding during this research.    
\end{acknowledgements}

\bibliographystyle{aa} 
\bibliography{NEP_MS} 


\begin{appendix} 

\section{Comparison of {\it Herschel} catalogues}
\label{appendix_herschel}

\begin{figure}[h!]
 \centering
  \includegraphics[width=\columnwidth]{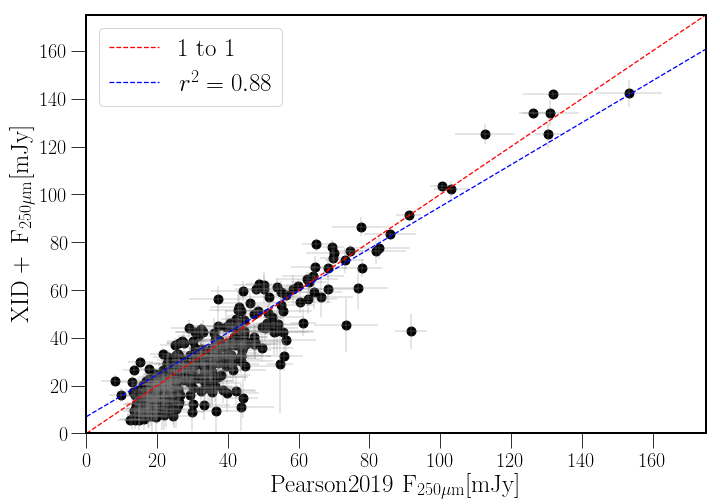} 
 \caption{ Comparison of the SPIRE 250$\mathrm{\, \mu m}$ band fluxes  from the catalogues of Pearson et al. in prep (x-axis) and HELP (y-axis), 
finding a good agreement between them. Also shown are the 1-to-1 line and   $\mathrm{r^2}$ regression line.} 
  \label{XID_herchel}
\end{figure}

As a validation of our SPIRE catalogue, the SPIRE fluxes from Pearson et al. (in prep.)  were compared with the those in the HELP catalogue \citep{Shirley2019}.  
We matched 370 SPIRE sources with sources in the HELP catalogue with high quality fluxes (the HELP catalogue also contains 2806 sources with low quality, unreliable fluxes) over an area of $\mathrm{\sim 2 deg^{2}}$. 
The results are shown in Figure  \ref{XID_herchel}. We find a very good agreement between the two catalogues, with  mean fluxes of ($\mathrm{35 \pm 1}$ mJy and of $\mathrm{32 \pm 1}$ mJy) from the Pearson catalogue and HELP catalogue respectively.

\section{High-z candidates} 
\label{appendix_highz}

Sources with detections in 6 or less  photometric bands were discarded from our analysis due to the low number of photometric bands (non of them have UV-optical data). 
This affects the spectral fit and the reliability in the calculation of stellar mass. 
However, they are interesting high-z sources that will be objects of further study. All the sources have at least one detection from two different telescopes. The sources are listed in Table \ref{List_rejecteddetections}. 

\begin{table}
 
 \begin{tabular}{|c|c|c|}
 \hline
 \textbf{ID} & \textbf{$\mathrm{z_{phot}}$} & \textbf{Photometric bands} \\ 
 \hline 
2113 	 & 4.8 $\pm$ 0.9      & 5  \\
2039	 & 4.7 $\pm$ 0.3      & 6  \\
1935	 & 4.7 $\pm$ 0.2      & 6  \\
2995     & 4.3 $\pm$ 0.7      & 5  \\
2267     & 4.2 $\pm$ 1.0      & 4  \\
4103	 & 4.2 $\pm$ 1.2      & 4  \\
2996     & 4.2 $\pm$ 1.2      & 6  \\
4398	 & 4.0 $\pm$ 1.3      & 4  \\
1729	 & 4.0 $\pm$ 1.0      & 4  \\
4438	 & 3.9 $\pm$ 1.2      & 5  \\
4115	 & 3.9 $\pm$ 1.1      & 6  \\
2077	 & 3.8 $\pm$ 1.1      & 5  \\
489	 & 3.8 $\pm$ 0.9      & 5  \\
2672	 & 3.8 $\pm$ 1.0      & 5  \\
NEP0049	 & 3.7 $\pm$ 0.9      & 6  \\
2732	 & 3.7 $\pm$ 0.9      & 5  \\
3889	 & 3.7 $\pm$ 1.1      & 4  \\
1343	 & 3.7 $\pm$ 1.0      & 5  \\
1219	 & 3.6 	$\pm$1.0      & 4  \\
1867	 & 3.6 	$\pm$0.9      & 5  \\
3461	 & 3.5 	$\pm$1.1      & 4  \\
2100	 & 3.5 	$\pm$0.9      & 5  \\
4804	 & 3.5 	$\pm$1.0      & 6  \\
241	 & 3.5 	$\pm$1.0      & 5  \\
2228	 & 3.4 $\pm$ 0.8      & 5  \\
1862	 & 3.4 $\pm$ 1.2      & 5  \\
3906	 & 3.3 $\pm$ 1.3      & 6  \\
2209	 & 3.3 $\pm$ 0.9      & 5  \\
3159	 & 3.3 	$\pm$1.2      & 5  \\
3617	 & 3.3 $\pm$ 1.2      & 5  \\
2674	 & 3.2 	$\pm$0.9      & 4  \\
2630	 & 3.2 $\pm$ 0.9      & 4  \\
4168	 & 3.2 $\pm$ 1.1      & 5  \\
1255	 & 3.2 	$\pm$1.1      & 4  \\
1296	 & 3.2 $\pm$ 0.9      & 5  \\
NEP0085	 & 3.2 	$\pm$0.9      & 5  \\
576	 & 3.1 	$\pm$0.7      & 5  \\
NEP0019  & 3.1 $\pm$ 0.8      & 4  \\
3318	 & 3.1 	$\pm$1.3      & 5  \\
3269	 & 3.1 $\pm$ 1.0      & 4  \\
4457	 & 3.1 	$\pm$0.8      & 4  \\
3645	 & 3.1 	$\pm$1.2      & 5  \\
3940     & 3.1 $\pm$ 1.0      & 5  \\
1417	 & 3.1 	$\pm$0.7      & 5  \\
NEP0027	 & 3.1 $\pm$ 1.0      & 4  \\
1100	 & 3.1 $\pm$ 0.8      & 4  \\
2830	 & 3.1 $\pm$ 0.7      & 5  \\
3211	 & 3.0 	$\pm$1.0      & 4  \\
2085	 & 3.0 $\pm$ 0.7      & 4  \\
1516	 & 3.0 	$\pm$1.1      & 5  \\
2687	 & 3.0 	$\pm$1.0      & 5  \\

\hline                               
    \end{tabular}                     
        \caption{List of the high-z candidates removed from the final list for only having a detection in 5 bands. The columns show the ID label, 
        the photometric redshift and the number of bands respectively.  }     
 \label{List_rejecteddetections}      
                          
 \end{table}

\end{appendix}

\end{document}